\documentclass[aps,pra,twocolumn,superscriptaddress,10pt,article,nofootinbib,showpacs]{revtex4-1}
\usepackage{amsfonts}
\usepackage{amsmath}
\usepackage{amsthm}
\usepackage{braket}
\usepackage{graphicx}
\usepackage{amssymb}
\usepackage{dsfont}
\usepackage{mathrsfs}
\usepackage{sidecap}
\usepackage[T1]{fontenc}
\usepackage[utf8]{inputenc}
\usepackage{float}
\usepackage{eufrak}
\usepackage{leftidx}

\newcommand{\otimesg}{\ensuremath{\otimes_{\mathfrak{g}}}}

\newcommand{\tr}{\ensuremath{\mathrm{tr}}}

\begin{document}

\title{Fermionic Matrix Product States and One-Dimensional Topological Phases} 

\author{Nick Bultinck}
\affiliation{Department of Physics and Astronomy, Ghent University, Krijgslaan 281, S9, 9000 Gent, Belgium}
\author{Dominic J. Williamson}
\affiliation{Vienna Center for Quantum Science and Technology, Faculty of Physics, University of Vienna, Boltzmanngasse 5, 1090 Vienna, Austria}
\author{Jutho Haegeman}
\affiliation{Department of Physics and Astronomy, Ghent University, Krijgslaan 281, S9, 9000 Gent, Belgium}
\author{Frank Verstraete}
\affiliation{Department of Physics and Astronomy, Ghent University, Krijgslaan 281, S9, 9000 Gent, Belgium}
\affiliation{Vienna Center for Quantum Science and Technology, Faculty of Physics, University of Vienna, Boltzmanngasse 5, 1090 Vienna, Austria}

\begin{abstract}
We develop the formalism of fermionic matrix product states (fMPS) and show how irreducible fMPS fall in two different classes, related to the different types of simple $\mathbb{Z}_2$ graded algebras, which are physically distinguished by the absence or presence of Majorana edge modes. The local structure of fMPS with Majorana edge modes also implies that there is always a two-fold degeneracy in the entanglement spectrum. Using the fMPS formalism we make explicit the correspondence between the $\mathbb{Z}_8$ classification of time-reversal invariant spinless superconductors and the modulo 8 periodicity in the representation theory of real Clifford algebras. Studying fMPS with general on-site unitary and anti-unitary symmetries allows us to define invariants that label symmetry-protected phases of interacting fermions. The behavior of these invariants under stacking of fMPS is derived, which reveals the group structure of such interacting phases.  We also consider spatial symmetries and show how the invariant phase factor in the partition function of reflection symmetric phases on an unorientable manifold appears in the fMPS framework. 
\end{abstract}

\maketitle
\tableofcontents

\section{Introduction} \label{sec:intro}

Tensor network states form a natural ansatz for the ground state of gapped local Hamiltonians because of their local entanglement structure. In one spatial dimension, it can even be proven that every ground state of such a Hamiltonian can be approximated by a tensor network state to arbitrary precision \cite{Hastings}. Such one-dimensional tensor networks, called matrix product states (MPS), thus capture the relevant subspace in which the low energy physics of gapped local systems takes place. For this reason, MPS are not only useful to describe model-dependent microscopic properties of quantum many-body systems, but also for universal properties associated to an entire family of Hamiltonians in the same quantum phase. It was realized that the universal properties that are stable under renormalization and manifest themselves in the far infrared field theory descriptions of the system are encoded in the entanglement properties of the MPS \cite{Li}. This insight has lead to a complete classification of bosonic quantum phases in one spatial dimension using tensor networks \cite{Pollmann1,Pollmann2,Chen1,Schuch2}. Although we will stick to one-dimensional systems in this work, we note for completeness that similar techniques applied to two-dimensional tensor network states have lead to a systematic understanding of the entanglement structure in non-chiral topological order and the properties of the associated superselection sectors \cite{Schuch,Buerschaper,Bultinck}. Also two-dimensional symmetry-protected phases \cite{Chen2,Williamson} and chiral phases \cite{Schuch3,Wahl,Dubail} can be described using the tensor network language. 

Extending the class of systems under consideration to also include fermionic degrees of freedom allows for a greater variety of quantum phases. For example, fermionic systems can have Majorana edge modes when defined on a lattice with boundary \cite{Kitaev1}. A complete classification of quantum phases for free fermion systems, known as the periodic table for topological insulators and superconductors, has been established through the use of tools such as Anderson localization and $K$-theory \cite{Kitaev2,Schnyder}. However, when interactions are taken into account the classification can change drastically \cite{Fidkowski3}. Recently it has been claimed that interacting symmetry-protected fermionic phases can be classified using cobordism theory \cite{Kapustin1,Freed}. In one spatial dimension, the classification of interacting fermionic systems was considered in \cite{Fidkowski2}. Building upon this work, we develop a tensor network approach towards symmetric fermionic phases. As a first step, in sections \ref{sec:super} and \ref{sec:fmps} we construct tensor networks that can carry fermionic degrees of freedom using the mathematical formalism of super vector spaces. We note that there already exist equivalent fermionic tensor network constructions in the literature using fermionic mode operators or Grassmann numbers~\cite{Corboz1,Kraus,Gu,Corboz2,Corboz3}, but for our purposes we find it more convenient to adopt the language of super vector spaces. Furthermore, these proposals are aimed towards two-dimensional tensor networks, as one-dimensional fermion systems are typically mapped to spin systems using the Jordan-Wigner transformation in numerical studies. However, the intrinsically fermionic  formalism has several advantages. For example, periodic or antiperiodic boundary conditions are automatically incorporated and  fermionic parent Hamiltonians can be constructed in a systematic way (see section \ref{sec:parent}). 

In section \ref{sec:fmps} we identify two distinct classes of irreducible fermionic matrix product states (fMPS), which are characterised by the presence or absence of Majorana zero modes at the ends of open chains.  Both classes or irreducible fMPS are shown to be the unique ground state or their associated fermionic parent Hamiltonian with periodic boundary conditions in section \ref{sec:parent}. The physical distinction between both types of fMPS has a clear signature in the algebraic structure of the local tensors constructing the fMPS and is related to the possible types of simple $\mathbb{Z}_2$ graded algebras over $\mathbb{C}$. This local tensor structure has a profound influence on the entanglement spectrum since it implies that Majorana edge modes are alway accompanied by a two-fold degeneracy of the Schmidt values, as we show in section \ref{sec:entanglement}. In section \ref{sec:timereversal} we include time-reversal symmetry, in the form of complex conjugation, and observe that the local structure of the fMPS tensors also provides an explicit link between the $\mathbb{Z}_8$ classification of time reversal symmetric spinless superconductors \cite{Fidkowski3} and the 8-fold periodicity in the representations of real Clifford algebras. Using the fundamental theorem of MPS in section \ref{sec:onsite} enables us to go beyond just time reversal and study general on-site unitary and anti-unitary symmetries by identifying invariants associated to the possible symmetric phases. Similar to the bosonic case, these invariants are obtained by studying the entanglement degrees of freedom in the ground state wave function. In this way we recover the classification derived in Ref.~\cite{Fidkowski2}, i.e. without Majorana edge modes a one-dimensional fermionic topological phase protected by the symmetry group $G$ is characterized by $H^2(G,U(1))$, while a fermionic symmetry protected phase with Majorana modes can only occur when $G = \tilde{G}\times \{I,P\}$ and corresponds to an element of $H^2(\tilde{G},U(1))$ and $H^1(\tilde{G},\mathbb{Z}_2)$. Sections \ref{sec:entanglement}, \ref{sec:timereversal}, \ref{sec:classification} and \ref{sec:antiunitary} are largely covered by Ref. \cite{Fidkowski2} but we revisit them using the fMPS formalism to make this work self-contained. In section \ref{sec:groupstructure} it is explained how the fMPS formalism is a natural framework for calculating the group structure of fermionic symmetric phases by studying the behavior of the invariants under stacking, which was not considered in Ref. \cite{Fidkowski2}. 

In section \ref{sec:reflection} we show that our method is not restricted to on-site symmetries and also recovers the $\mathbb{Z}_8$ group structure of reflection symmetric phases. This result relies heavily on the intrinsically fermionic nature of our formalism. To obtain the correct group structure we use the recently proposed method of partial reflection \cite{Shapourian,Shiozaki} and show that it fits comfortably within the tensor network language.  

Since the fMPS formalism is an extension of bosonic matrix product states, we first review some basic facts about them in section \ref{bosonicMPS}.

\section{Bosonic matrix product states}\label{bosonicMPS}
A matrix product state (MPS) is defined in the Hilbert space $\mathcal{H}^1\otimes \mathcal{H}^2\otimes\dots\otimes\mathcal{H}^N$ of a one dimensional lattice of $N$ finite-dimensional bosonic degrees of freedom. A general state is defined by the array of coefficients $C^{i_1,i_2,\ldots,i_N}$ with respect to the product basis $\ket{i_1} \otimes \ket{i_2} \otimes \cdots \otimes \ket{i_N}$. To construct a MPS one associates a rank three array $A^{i}[j]_{\alpha\beta}$ to every site $j$, where the index $i$ is associated with the basis $\ket{i}$ of the local physical Hilbert space $\mathcal{H}^j$ and the so-called virtual indices $\alpha$ and $\beta$ are of dimension $D_{j}$ and $D_{j+1}$ respectively. We also denote by $A^i[j]$ the $D_{j} \times D_{j+1}$ matrix obtained by fixing the index $i$. The MPS is then defined as
\begin{equation}\label{MPS}
|\psi\rangle=\sum_{\{i_j\}}\text{tr}\left(A^{i_1}[1]A^{i_2}[2]\dots A^{i_N}[N] \right)|i_1\rangle |i_2\rangle\dots |i_N\rangle\, .
\end{equation}
When all local Hilbert spaces $\mathcal{H}^j\equiv \mathbb{C}^d$ with constant physical dimension $d$, we can associate the same $A^i_{\alpha\beta}$ to every site in order to obtain translation invariant states \footnote{A translation invariant MPS is not necessarily of this form and can for example have an additional matrix $B$ in the trace that commutes with all of the $A^i$. For an irreducible MPS (see below), such a $B$ naturally satisfies $B\sim 1$, but in the case of fermionic MPS, a non-trivial $B$ can arise.}. The dimension $D$ of the virtual indices is than independent of the links and is referred to as the bond dimension. From the definition \eqref{MPS} it is clear that the resulting MPS is invariant under a similarity transformation of the matrices $A^i$, i.e.\ if we replace the $d$ matrices $A^i$ with $XA^iX^{-1}$ then the resulting state $|\psi\rangle$ remains the same. For this reason we call such a transformation of the tensor $A$ a gauge transformation.

MPS of this form can be brought into a canonical form \cite{Fannes,PerezGarcia}. In discussing this canonical form we will follow the presentation of Ref.~\cite{Cirac}. Let us consider the situation where we decompose the identity on the virtual indices as a sum of two projectors $P$ and $P^\perp=1-P$, where $P$ corresponds to the orthogonal projector onto an invariant subspace of the matrices $A^i$, i.e.
\begin{align}
A^iP &= PA^iP\\
P^{\perp} A^i &= P^{\perp} A^iP^{\perp}\, .
\end{align}
Using these relations one sees that the MPS
\begin{equation}
|\psi\rangle=\sum_{\{i_j\}}\text{tr}\left(A^{i_1}A^{i_2}\dots A^{i_N} \right)|i_1\rangle |i_2\rangle\dots |i_N\rangle
\end{equation}
can equivalently be constructed by replacing the matrices $A^i$ with $\hat{A}^i$, where
\begin{equation}
\hat{A}^i = PA^iP + P^{\perp} A^iP^{\perp}  \equiv A^i_1\oplus A^i_2\, .
\end{equation}
We can now apply the same argument to further decompose the $A^i_1$ and $A^i_2$ blocks under the presence of possible invariant subspaces. After a finite number of iterations we can write the matrices as
\begin{equation}
\hat{A}^i = \bigoplus_{k=1}^r A^i_k\, ,
\end{equation}
where none of the blocks $A^i_k$ contains an invariant subspace. This form of the matrices implies that we can decompose $|\psi\rangle$ as a sum of multiple MPS: $|\psi\rangle = \sum_{k=1}^r |\psi\rangle_k$, where $|\psi\rangle_k$ is a MPS constructed from the matrices $A^i_k$. The existence of the canonical form implies that without loss of generality we can restrict to matrices $\hat{A}^i$ that span a semisimple algebra. If the $A^i$ span a simple $D\times D$ matrix algebra we see that the corresponding MPS is irreducible, i.e. it cannot be decomposed as a sum of multiple MPS. This class of irreducible MPS is commonly refered to as injective MPS because they satisfy the property that there exists a $p \in \mathbb{N}$, such that for all $q > p$ we have that $\left(A^{i_1}A^{i_2}\dots A^{i_q}\right)_{\alpha\beta}$, interpreted as a map from $\alpha,\beta$ to $i_1,i_2,\dots,i_q$, is injective.

To every such MPS we can also associate a so-called parent Hamiltonian \cite{Fannes}. It consists of a sum of local terms acting on $q>p$ neighbouring sites. To define the local terms we again consider the map $\left(A^{i_1}A^{i_2}\dots A^{i_q}\right)_{\alpha\beta}$. The local Hamiltonian terms are then projectors onto the space orthogonal to the image of this map. Because the image is $D^2$ dimensional and the total physical space corresponding to $q$ site is $d^q$, this space orthogonal to the image is alway non-zero for large enough $q$. By construction, the parent Hamiltonian of a MPS is frustration free and has the MPS as its ground state. If the MPS is irreducible, one can even show that the parent Hamiltonian is gapped and that its ground state is unique. If the MPS is reducible, i.e. $|\psi\rangle = \sum_{k=1}^r |\psi\rangle_k$, then every state $|\psi\rangle_k$ is by itself a ground state of the parent Hamiltonian.

\section{Super vector spaces}\label{sec:super}

To construct fermionic matrix product states we will make use of super vector spaces. In this section we introduce the relevant concepts and present the notation to be used in following sections. A super vector space $V$ has a natural direct sum decomposition
\begin{equation}
V = V^0\oplus V^1\, ,
\end{equation}
where we refer to vectors in $V^0 (V^1)$ as even (odd) parity vectors. Vectors that have a definite parity are called homogeneous. We denote the parity of a homogeneous basis state $|i\rangle$ by $|i|\in \{0,1\}$. The tensor product of two homogeneous vectors $|i\rangle$ and $|j\rangle$ is again a homogeneous vector, and its parity is given by $|i|+|j|$ mod 2. In other words, $V$ and the associated operation of taking tensor products is $\mathbb{Z}_2$ graded. We will denote the graded tensor product as
\begin{equation}
|i\rangle \otimesg|j\rangle \in V\otimesg V\, .
\end{equation}
The key relation between super vector spaces and fermionic degrees of freedom is the following canonical graded tensor product isomorphism
\begin{align}\label{reordering}
\mathcal{F}:& V\otimesg W \rightarrow W\otimesg V \nonumber \\
 & |i\rangle\otimesg|j\rangle \rightarrow (-1)^{|i||j|}|j\rangle\otimesg|i\rangle
\end{align}
The canonical isomorphism $\mathcal{F}$ is the crucial ingredient of super vector spaces and it shows why even (odd) parity vectors can be interpreted as having even (odd) fermion number. 

The dual space $V^*$, defined via its linear action on $V$, has a canonical basis $\bra{i}$ satisfying $\braket{i|j} = \delta_{i,j}$ that inherits the $\mathbb{Z}_2$ grading of $V$. In particular, the definition of $\mathcal{F}$ has a natural extension to also include the action on the relevant dual spaces. For example, if we replace $V$ with its dual space $V^*$, then $\mathcal{F}$ acts as
\begin{align}
\mathcal{F}:& V^*\otimesg W \rightarrow W\otimesg V^* \nonumber \\
 & \langle i| \otimesg|j\rangle \rightarrow (-1)^{|i||j|}|j\rangle\otimesg\langle i|\, ,
\end{align}
and similarly for the other cases $V\otimesg W^*$ and $V^*\otimesg W^*$. We will often refer to $\mathcal{F}$ as fermionic reordering and use the notation $\mathcal{F}$ very loosely, i.e. every isomorphism between graded tensor products of super vector spaces that corresponds to multiple applications of $\mathcal{F}$ as defined above will actually be denoted with the same symbol, and clearly the precise order of such subsequent applications is irrelevant.

A tensor in $V^*\otimesg V$ can be mapped to $\mathbb{C}$ by using the natural action of the dual space. We denote this homomorphism as a contraction $\mathcal{C}$
\begin{equation}\label{contraction}
\mathcal{C}:V^*\otimesg V \rightarrow \mathbb{C}: \langle \psi |\otimes_{\mathfrak{g}} |\phi\rangle \rightarrow \langle \psi|\phi\rangle\, ,
\end{equation}
and in particular $\mathcal{C}(\bra{j} \otimesg \ket{i}) = \delta_{i,j}$. Applying the contraction of $V^*$ and $V$ in a more general tensor product involving several super vector spaces requires that we first apply $\mathcal{F}$ so as to isolate $V^*\otimesg V$. Still denoting this combined reordering and contraction as $\mathcal{C}$, we obtain for example the famous supertrace
\begin{equation}
\mathcal{C}(\ket{i} \otimesg \bra{j}) = (-1)^{|i||j|} \mathcal{C}(\bra{j} \otimesg \ket{i}) = (-1)^{|i|} \delta_{i,j}.	
\end{equation}
To obtain the normal trace, we need to include the fermion parity operator $\sum_{k} (-1)^{|k|} \ket{k} \otimesg \bra{k}$ so that we indeed obtain
\begin{equation}
\mathcal{C}(\sum_{k} (-1)^{|k|} \ket{k} \otimesg \bra{k} \otimesg\ket{i} \otimesg \bra{j}) = (-1)^{|i|} \mathcal{C}(\ket{i} \otimesg \bra{j}) = \delta_{i,j}.	
\end{equation}
The contraction map also defines the canonical isomorphism $(V\otimesg W)^* \simeq W^* \otimesg V^*$, as indeed we have $\mathcal{C}(\bra{j'}\otimesg \bra{i'} \otimesg \ket{i} \otimesg \ket{j} = \delta_{i',i} \delta_{j',j}$.

The grading of the super vector space $V$ carries over to the algebra of (anti-)linear operators acting on $V$, as linear operators on $V$ are naturally represented as tensors in $V\otimesg V^*$ as
\begin{equation}\label{M}
\mathsf{M} = \sum_{i,j}M_{ij}|i\rangle\otimes_{\mathfrak{g}}\langle j| \in V\otimes_{\mathfrak{g}}V^*\, .
\end{equation}
The algebra of operators on $V$ thus becomes a superalgebra, whose homogenous elements are represented by tensors $\mathsf{M}$ with a well defined parity, henceforth denoted as $|\mathsf{M}|$, i.e.\ $|i|+ |j| \mod 2 = |\mathsf{M}|$ is equal for all contributions to the sum \eqref{M}. The algebra multiplication rule $\cdot$ is obtained by applying the contraction map $\mathcal{C}$, which involves the fermionic reordering $\mathcal{F}$. For the algebra of operators on $V\otimesg W$, this results in the multiplication rule
\begin{equation}
\left( \mathsf{M} \otimes_{\mathfrak{g}} \mathsf{N}\right)\cdot \left( \mathsf{O}\otimes_{\mathfrak{g}}\mathsf{P}\right) = (-1)^{|\mathsf{N}||\mathsf{O}|} (\mathsf{M}\cdot \mathsf{O})\otimes_{\mathfrak{g}}(\mathsf{N}\cdot \mathsf{P})\, ,
\end{equation}
with $\mathsf{M}, \mathsf{O} \in V\otimesg V^*$ and $\mathsf{N}, \mathsf{P} \in W\otimesg W^*$.

\section{Fermionic matrix product states}\label{sec:fmps}

In this section we introduce the general formalism of fermionic matrix product states (fMPS). We obtain two distinct classes, one leading to even parity states with periodic boundary conditions, and one to odd parity states. But firstly, we introduce the general notion of fermionic tensor networks using super vector spaces.

\subsection{Fermionic tensor networks}
We start by providing a more abstract definition of the bosonic MPS. We therefore promote the rank three arrays $A^i_{\alpha,\beta}$ to tensors
\begin{equation}
\mathsf{A}[j] = \sum_{i,\alpha,\beta} A[j]^i_{\alpha,\beta} |\alpha)_{j-1}\otimes |i\rangle_j \otimes (\beta|_j \in V^j\otimes \mathcal{H}^j \otimes (V^{j+1})^*\,,
\end{equation}
where round bras and kets correspond to the basis of the virtual spaces $V^j \simeq \mathbb{C}^{D_j}$ (and its duals). The MPS $\ket{\psi}$ from Eq.~\eqref{MPS} is then obtained as
\begin{equation}
\ket{\psi} = \mathcal{C}_{v}(\mathsf{A}[1] \otimes \mathsf{A}[2] \otimes \cdots \otimes \mathsf{A}[N])
\end{equation}
where $\mathcal{C}_{v}$ denotes the contraction of all the virtual indices and the different tensor product orders are trivially isomorphic in the bosonic case (reordering does not introduce signs).

A natural definition of a fMPS follows from this construction by starting from tensors $\mathsf{A}[j] \in V^j \otimesg \mathcal{H}^j \otimesg (V^{j+1})^*$ where both the physical Hilbert space $\mathcal{H}^j$ and the virtual spaces $V^j$ are super vector spaces. The fMPS is then obtained as
\begin{equation}
\ket{\psi} = \mathcal{C}_{v}(\mathsf{A}[1] \otimesg \mathsf{A}[2] \otimesg \cdots \otimesg \mathsf{A}[N])
\end{equation}
where the contraction $\mathcal{C}_v$ over the virtual indices now involves the fermionic reordering isomorphism $\mathcal{F}$. This construction extends to fermionic tensor networks in general. If all of the individual tensors have a well defined parity $|\mathsf{A}[j]|$, then so does the resulting state and a different initial order of the tensors in the graded tensor product will at most result in a global sign difference of the state. In particular, if at most one of the tensors is odd, the definition of the fermionic tensor network is independent of the order of the individual tensors. As the ability to manipulate tensor networks locally is of paramount importance both for numerical as well as theoretical applications, we will always impose this constraint as a consistency condition.

As an illustration, let us define following fermionic tensors
\begin{align*}
C &= \sum_{\alpha\beta\gamma}C_{\alpha\beta\gamma}|\alpha)|\beta)(\gamma| \\
D &= \sum_{\lambda\kappa}D_{\lambda\kappa}|\lambda)(\kappa|
\end{align*}
Suppose we wish to contract the $\beta$ index of $C$ with the $\kappa$ index of $D$. As explained above, we first take the graded tensor product of $C$ and $D$:
\begin{displaymath}
C\otimes_{\mathfrak{g}} D = \sum_{\alpha\beta\gamma\lambda\kappa}C_{\alpha\beta\gamma}D_{\lambda\kappa}|\alpha)|\beta)(\gamma|\otimes_{\mathfrak{g}}|\lambda)(\kappa|\, .
\end{displaymath}
Next, we bring the $\kappa$ bra next to the $\beta$ ket using fermionic reordering
\begin{displaymath}
\mathcal{F}(C\otimes_{\mathfrak{g}} D) = \sum_{\alpha\beta\gamma\lambda\kappa}C_{\alpha\beta\gamma}D_{\lambda\kappa}(-1)^{|\kappa|(|\lambda|+|\gamma|+|\beta|)}|\alpha)(\kappa|\,|\beta)(\gamma|\,|\lambda)\, .
\end{displaymath}
If the tensors $C$ and $D$ are even, this is equivalent to
\begin{displaymath}
\mathcal{F}(C\otimes_{\mathfrak{g}} D) = \sum_{\alpha\beta\gamma\lambda\kappa}C_{\alpha\beta\gamma}D_{\lambda\kappa}(-1)^{|\kappa|+|\kappa||\alpha|}|\alpha)(\kappa|\,|\beta)(\gamma|\,|\lambda)\, .
\end{displaymath}
Now we apply the contraction to obtain the final tensor:
\begin{displaymath}
F \equiv \sum_{\alpha\gamma\lambda}\left(\sum_\beta C_{\alpha\beta\gamma}D_{\lambda\beta}(-1)^{|\beta|+|\beta||\alpha|}\right)|\alpha)(\gamma|\,|\lambda)\, .
\end{displaymath}
Note that also in the definition of the contracted tensor $F$ we have to include an internal ordering. Different internal orderings give additional minus signs to the tensor components. It therefore only makes sense to compare tensors that have the same internal ordering. Let us now use this formalism to study the two different classes of fMPS.

\subsection{Even parity states}\label{evenfMPS}
To build translation invariant fMPS on chains of every length, we start from even tensors $\mathsf{A} = \sum_{i\alpha\beta} A^i_{\alpha,\beta} |\alpha)\otimesg |i\rangle \otimesg (\beta|$. As in the bosonic case, we will denote by $A^i$ the matrices obtained by fixing the superscript $i$ in the array $A^i_{\alpha,\beta}$. 
Evenness of the tensors requires these matrices to take the following form
\begin{equation}\label{evenalgebra}
\begin{split}
A^i &= \left(\begin{matrix} B^i & 0 \\ 0 & C^i  \end{matrix} \right)\;\;\;\;\text{ if } |i| = 0 \\
A^i &= \left(\begin{matrix} 0 & D^i \\ F^i & 0  \end{matrix} \right)\;\;\;\;\text{ if } |i| = 1\, ,
\end{split}
\end{equation}
in a standard basis of the virtual space where $|\alpha|=0$ for $\alpha=1,\ldots,D_e$ and $|\alpha|=1$ for $\alpha = D_e+1,\ldots,D_e+D_o = D$. Note that $D_e$ and $D_o$ can be different.
\begin{widetext}
By the chosen internal order of the tensors $\mathsf{A}$, the contraction of the virtual indices on the bonds $1$ to $N-1$ is trivial and gives rise to 
\begin{displaymath}
\ket{\psi}_{e} =  \mathcal{C}_N \left( \sum_{\{i\}}\sum_{\alpha\beta}\left(A^{i_1}A^{i_2}\dots A^{i_N}\right)_{\alpha\beta}|\alpha)_N |i_1\rangle |i_2\rangle \dots |i_N\rangle(\beta|_N\right)\, .
\end{displaymath}
where no minus signs have been generated and only the virtual index on bond $N$ remains to be contracted. Using that all tensors are even, we can apply fermionic reordering to obtain
\begin{displaymath}
\ket{\psi}_{e} =  \mathcal{C}_N\left( \sum_{\{i\}}\sum_{\alpha\beta}\left(A^{i_1}A^{i_2}\dots A^{i_N}\right)_{\alpha\beta}(-1)^{|\beta|}(\beta|\,|\alpha) |i_1\rangle |i_2\rangle \dots |i_N\rangle\right)\, ,
\end{displaymath}
\end{widetext}
where we can apply the contraction trivially in order to obtain
\begin{equation}
|\psi\rangle_{e} = \sum_{\{i\}} \text{tr}\left(\mathcal{P} A^{i_1}A^{i_2}\dots A^{i_N} \right) |i_1\rangle |i_2\rangle \dots |i_N\rangle\,.
\end{equation}
Here, we introduced the parity matrix $\mathcal{P}$
\begin{equation}\label{eq:parity}
\mathcal{P} = \left( \begin{matrix} \mathds{1} & 0 \\ 0 &  -\mathds{1}  \end{matrix}\right)\,
\end{equation}
which has the defining property
\begin{equation}\label{eq:parityproperty}
\mathcal{P}A^{i} = \left( -1\right)^{|i|}A^{i}\mathcal{P}\, ,
\end{equation}
because we started from even tensors $\mathsf{A}$. The resulting fMPS $|\psi\rangle_e$, being the contraction of these even tensors, has even fermion parity, as indicated by the subscript. One sees that the coefficients of the fMPS satisfy
\begin{displaymath}
\text{tr}\left(\mathcal{P} A_e^{i_1}A_e^{i_2}\dots A_e^{i_N}\right) = \left(-1 \right)^{|i_1|}\text{tr}\left(\mathcal{P} A_e^{i_2}A_e^{i_3}\dots A_e^{i_1}\right)\, ,
\end{displaymath}
which is indeed the correct behaviour for translationally invariant fermionic states with even fermion parity, as shown in appendix \ref{app:translation}.

\subsection{Odd parity states}\label{sec:oddparity}
Because the fMPS tensors $\mathsf{A}$ are even, a fMPS with odd fermion parity is obtained by adding one additional tensor with odd parity and no physical component to the tensor network. We choose $\mathsf{Y} = \sum_{\alpha,\beta} Y_{\alpha,\beta} |\alpha)_N \otimesg (\beta|_1$ with $Y_{\alpha,\beta}=0$ if $|\alpha|+|\beta|\mod 2 = 0 $.
Evaluating $\ket{\psi}_o = \mathcal{C}_v\left(\mathsf{Y} \otimesg \mathsf{A}[1] \otimesg \mathsf{A}[2]\otimesg \cdots \otimesg \mathsf{A}[N]\right)$ using the same steps as in the previous subsection results in
\begin{equation}\label{oddfMPS}
|\psi\rangle_{o} = \sum_{\{i\}} \text{tr}\left(Y A^{i_1}A^{i_2}\dots A^{i_N} \right) |i_1\rangle |i_2\rangle \dots |i_N\rangle\,.
\end{equation}
If we require this state to be invariant under translations then, as shown in appendix \ref{app:translation}, it should hold that
\begin{displaymath}
\text{tr}\left(Y A^{i_1}A^{i_2}\dots A^{i_N} \right) =  \text{tr}\left(Y A^{i_2}A^{i_3}\dots A^{i_1} \right)\, ,
\end{displaymath}
implying that $Y$ has to commute with all $A^i$. To satisfy these requirements we choose $D_e = D_o$ and take the tensors to be of the form
\begin{subequations}\label{oddalgebra}
\begin{equation}
\begin{split}
A^i & = \left(\begin{matrix} B^i & 0 \\ 0 & B^i \end{matrix}\right) = \mathds{1} \otimes B^i\;\;\;\;\;\;\text{ if } |i| = 0 \\
A^i & = \left(\begin{matrix} 0 & B^i \\ -B^i & 0 \end{matrix}\right) = y \otimes B^i \;\;\;\;\text{ if } |i| = 1\, ,
\end{split}
\end{equation}
with
\begin{equation}
y = \left(\begin{matrix}0 & 1 \\ -1 & 0 \end{matrix}\right)\, .
\end{equation}
The odd matrix commuting with $A^i$ is then simply
\begin{equation}
Y = \left(\begin{matrix} 0 & \mathds{1} \\ -\mathds{1} & 0 \end{matrix} \right) = y\otimes \mathds{1}\, .
\end{equation}
\end{subequations}
We will comment on the generality of this choice in the next section.

Let us now look at what happens when we define the fMPS with odd fermion parity on a chain with open boundary conditions. We will do this by looking at a particular example, namely the Kitaev chain for spinless fermions, which is described by the Hamiltonian \cite{Kitaev1}
\begin{equation} \label{KitaevChain}
H_{\text{Kitaev}} = -i\sum_{j = 1}^N\gamma_{2j}\gamma_{2j+1}\, ,
\end{equation}
where
\begin{eqnarray}
\gamma_{2j -1} & = &  -i(a_j - a^\dagger_j) \\
\gamma_{2j}  & = & a_j + a^\dagger_j
\end{eqnarray}
are Majorana operators satisfying $\gamma_j^\dagger = \gamma_j$ and $\{\gamma_j,\gamma_k\} = 2\delta_{jk}$. We can easily obtain the exact fMPS description of the ground state with periodic boundary conditions by applying the projectors
\begin{equation}\label{kitaevprojector}
P_j = \frac{1}{2}\left(\mathds{1} + i\gamma_{2j}\gamma_{2j+1} \right) = \frac{1}{2}\left(\mathds{1} - (a^\dagger_j + a_j)(a^\dagger_{j+1} - a_{j+1}) \right)
\end{equation}
to an arbitrary state with odd fermion parity (acting with these projectors on an even parity state gives zero). The matrices of the fMPS ground state are:
\begin{eqnarray}\label{Kitaevchain}
A^0 & = & \left(\begin{matrix}1 & 0 \\ 0 & 1 \end{matrix}\right) \\
A^1  & = & Y = y =  \left(\begin{matrix}0 & 1 \\ -1 & 0 \end{matrix}\right) \, ,
\end{eqnarray}
which is indeed a special case of the general structure given in equations \eqref{oddalgebra}. Starting from the expression
\begin{equation}
\sum_{\{i\}}\left(A^{i_1}A^{i_2}\dots A^{i_N} \right)_{\alpha\beta}|\alpha) |i_1\rangle |i_2\rangle\dots |i_N\rangle (\beta|
\end{equation}
we can now obtain four ground states on a chain with open boundary conditions by closing the virtual indices at the boundaries with either $(0|$ ($|0)$) or $(1|$ ($|1)$). However, because of the special structure of the tensor, the two states with even fermion parity, obtained by closing the virtual indices diagonally with either $(0|\otimes_{\mathfrak{g}}|0)$ or $(1|\otimes_{\mathfrak{g}}|1)$ are equal. Also the two odd parity states obtained by closing off-diagonally with $(0|\otimes_{\mathfrak{g}}|1)$ or $(1|\otimes_{\mathfrak{g}}|0)$ are equal up to a minus sign. So on a chain with open boundary conditions we have only two different ground states, one with even and one with odd parity. The information about which ground state we pick is shared between the two edges; i.e. it is encoded in the way the fMPS was closed, either diagonally or off-diagonally. Because there is no local way to detect this difference, there is no local term that can be added to the Hamiltonian to split the degeneracy for large system sizes. This is the fMPS manifestation of the appearance of Majorana edge modes \cite{Kitaev1}.

\subsection{Irreducibility}\label{sec:irreducibility}
At this point the structure of the tensors we used to construct fMPS with odd parity \eqref{oddalgebra} can be seen as a special case of the structure of tensors for even parity fMPS \eqref{evenalgebra}. Furthermore, interpreting the matrices $A^i$ as those specifying a bosonic MPS, the existence of a matrix $Y$ commuting with all $A^i$ in the case of the odd parity states would point towards a reducible representation and hence symmetry breaking. We therefore need to redevelop the concept of irreducibility for fMPS from the ground up, using the notion of invariant subspaces. We thus start from the matrices $A^i$ with the general structure of Eq.~\eqref{evenfMPS}, whose defining signature is the existence of the parity matrix $\mathcal{P}=\mathcal{P}^{-1}$ satisfying Eq.~\eqref{eq:parityproperty}. If the matrices $A^i$ have a non-trivial invariant subspace with corresponding orthogonal projector $P$ satisfying
\begin{displaymath}
A^i P = P A^i P
\end{displaymath}
then necessarily
\begin{displaymath}
\mathcal{P} A^i \mathcal{P} P = P \mathcal{P} A^i \mathcal{P} P
\end{displaymath}
and thus
\begin{displaymath}
 A^i \mathcal{P} P \mathcal{P} = \mathcal{P} P \mathcal{P} A^i \mathcal{P} P \mathcal{P}
\end{displaymath}
so that $Q = \mathcal{P} P \mathcal{P}$ is also an orthogonal projector onto an invariant subspace. If $P$ was already associated to an irreducible invariant subspace (containing no smaller non-trivial invariant subspaces), then either $Q=P$ or $PQ=QP=0$. Otherwise, the intersection of the invariant subspaces of $Q$ and $P$ would be an invariant subspace of its own, thus leading to a contradiction.

The case $Q=P$ corresponds to $[P,\mathcal{P}]=0$ and allows one to decompose the invariant subspace (and its orthogonal complement) into an even and odd part using the projectors $P_{\pm} = (1\pm \mathcal{P})/2$ as $P = P_+ P P_+ + P_- P P_-$. Consequently, we can replace $A^i$ by $\hat{A}^i = PA^iP + P^{\perp} A^i P^{\perp}$ where both $PA^iP$ and $P^{\perp} A^i P^{\perp}$ individually have the structure of Eq.~\eqref{eq:parityproperty} and thus correspond to even fermionic tensors specifying a fMPS.

If the matrices $A^i$ have no more non-trivial invariant subspaces satisfying $[P,\mathcal{P}] = 0$, we can try to reduce them further using invariant subspaces where $P Q = QP = 0$. Note, firstly, that $P+Q$ is an invariant subspace projector that does commute with $\mathcal{P}$ and thus, by the previous assumption, is equal to the identity. This imposes the following structure on $P$ and $Q=P^{\perp}=1-P=\mathcal{P} P \mathcal{P}$
\begin{align}
P&=\frac{1}{2} \begin{bmatrix}
\mathds{1} & U\\
U^\dagger & \mathds{1}
\end{bmatrix},&Q&=\frac{1}{2} \begin{bmatrix}
\mathds{1} & -U\\
-U^\dagger & \mathds{1}
\end{bmatrix},
\end{align}
where idempotence requires $U U^\dagger = U^\dagger U = \mathds{1}$. Hence, this is only possible if $D_e=D_o$ and $U$ is a unitary matrix. This implies the matrices $A^i$ are of the following form
\begin{equation}\label{eq:oddalgebrageneric}
\begin{split}
A^i &= \left(\begin{matrix} B^i & 0 \\ 0 & U^\dagger B^i U \end{matrix}\right) \;\;\;\;\text{ if } |i| = 0 \\
A^i &= \left(\begin{matrix} 0 & B^i \\ U^\dagger B^i U^\dagger & 0 \end{matrix}\right) \;\;\;\;\text{ if } |i| = 1\, .
\end{split}
\end{equation}
An even-parity gauge transform of the form $\mathds{1} \oplus \mathrm{i}U$ will map this to the standard form of Eq.~\eqref{oddalgebra}, which we will employ for the remainder of the discussion.

So why, despite the presence of non-trivial invariant subspaces, is this fMPS irreducible? So far, our irreducibility discussion has not distinghuished between fermionic MPS or bosonic MPS with a $\mathbb{Z}_2$ symmetry. In the bosonic case, the latter form can be further reduced using $P$ and $P^{\perp}$ into two MPS, which will be irreducible if the $B^i$ span a simple matrix algebra of dimension $D_e=D_o=D/2$. These individual MPS break the $\mathbb{Z}_2$ symmetry. The even and odd superposition on the periodic chain are obtained by closing the MPS  with either the identity or with $Y$. In the fermionic case, the new matrices obtained by reducing $A^i$ with $P$ and $P^{\perp}$ do not make sense as fermionic tensors, as they do not have well defined parity. Indeed, fermion parity cannot be broken. On the periodic chain, we can again try to create the symmetric and antisymmetric linear combinations corresponding to even and odd fermion parity. To close the fMPS, we also have to start at the level of the fermionic tensors, where we can thus try to add either the identity or the non-trivial element $\mathsf{Y}$, giving rise to the two states $\mathcal{C}_v( \mathsf{A}^{\otimesg N})$ and $\mathcal{C}_v( \mathsf{Y}\otimes \mathsf{A}^{\otimesg N})$. However, unlike in the bosonic case, upon applying the fermionic contraction and reordering, the first state evaluates to zero because an extra factor $\mathcal{P}$ is introduced at the level of the matrices, as discussed in Section~\ref{evenfMPS}. So only the odd parity state survives as a translation invariant state.

We thus want to conclude that fermionic tensors $\mathsf{A}$ associated with the matrices $A^i = y^{|i|}\otimes B^i$ define an irreducible fMPS if the $B^i$ span a simple $D/2 \times D/2$ dimensional matrix algebra. However, this is not sufficient as there is an extra condition hidden in the fact that $A^i$ also doesn't have an invariant subspace that commutes with $\mathcal{P}$. Suppose indeed that $A^i= y^{|i|}\otimes B^i$ has a non-trivial invariant subspace $A^i P = P A^i P$ with $P$ of the form
\begin{equation}
P = \begin{bmatrix}
P_e & 0\\0 & P_o
\end{bmatrix}\, ,
\end{equation}
and $P_o = \mathds{1}_{D/2}-P_e$. The condition $A^iP = PA^iP$ then implies that
\begin{align}\label{invariantP}
B^i P_e &= P_e B^i P_e, \forall |i|=0,& B^i P_o &= P_o B^i P_o, \forall |i|=0,\nonumber\\
B^i P_e &= P_o B^i P_e, \forall |i|=1,& B^i P_o &= P_e B^i P_o, \forall |i|=1.
\end{align}
This type of invariant subspace is consistent with the requirement that the matrices $B^i$ span a simple $D/2\times D/2$ matrix algebra.  However, when properties \eqref{invariantP} hold we can conclude that the even subalgebra, spanned by $B^{i_1}\dots B^{i_p}$ with $p \in \mathbb{N}$ and $\sum_{j=1}^p |i_j| = 0$ mod 2 has $P_e$ and $P_o$ as non-trivial invariant subspaces. A sufficient condition for the odd parity fMPS to be irreducible is thus that the even subalgebra spanned by $y^{|i|}\otimes B^i$ should be a simple $D/2\times D/2$ matrix algebra. This is also a necessary condition, as an irreducible algebra $B^i$ with a reducible even subalgebra automatically leads to the existence of a $P_e$ and $P_o = P_e^{\perp}$ and thus to an invariant subspace projector $P$ satisfying $[P,\mathcal{P}]$ for the $A^i$. The physical reason for excluding this case is that the above structure of the $B^i$s, in combination with the fact that the resulting state has an odd fermion parity and thus an odd number of $B^i$ factors with $|i|=1$ automatically makes the state zero, as can readily be verified.

In summary, an fMPS is irreducible in the following two cases. In the standard basis where $\mathcal{P}=\mathds{1}_{D_e} \oplus -\mathds{1}_{D_o}$, the matrices $A^i$ either
\begin{itemize}
	\item take the form of Eq.~\eqref{evenalgebra} and span a simple $D\times D$ matrix algebra; the center is trivial and resulting translation invariant fMPS on the periodic chain have even fermion parity. 
	\item take the form of Eq.~\eqref{eq:oddalgebrageneric}, which can be gauge transformed into the canonical form of Eq.~\eqref{oddalgebra} where $A^i=y^{|i|} \otimes B^i$, and the even subalgebra of the $B^i$ is a simple $D/2\times D/2$ matrix algebra. The resulting translation invariant fMPS on a periodic chain has odd fermion parity.
\end{itemize}
The above two notions of irreducibility for even and odd parity fMPS correspond to the two possibilities for simple $\mathbb{Z}_2$ graded algebras \cite{Fidkowski2,Wall}. An even simple $\mathbb{Z}_2$ graded algebra is simple as an ungraded algebra, which implies that its center consists only of multiples of the identity. An odd simple $\mathbb{Z}_2$ graded algebra $\mathcal{A} = \mathcal{A}_0\oplus \mathcal{A}_1$, where $\mathcal{A}_0$ consists of the even parity elements and $\mathcal{A}_1$ of the odd parity elements, has the property that $\mathcal{A}_0$ is simple and $\mathcal{A}_1 = Y\mathcal{A}_0$, with $Y$ an odd element satisfying $Y^2 \propto \mathds{1}$. The graded center of an odd simple algebra consists only of multiples of $\mathds{1}$ and $Y$. So we see that the bosonic statement of irreducibility:
\begin{center}
\emph{A MPS is irreducible $\Leftrightarrow$ $A^i$ span a simple algebra}
\end{center}
has an elegant generalization to the fermionic case:
\begin{center}
\emph{A fMPS is irreducible $\Leftrightarrow$ $A^i$ span a simple $\mathbb{Z}_2$ graded algebra}
\end{center}
The two types of simple $\mathbb{Z}_2$ graded algebras, which are called even and odd type, correspond to fMPS with even or odd fermion parity respectively under periodic boundary conditions. For this reason we henceforth refer to these two types of irreducible fMPS as even algebra and odd algebra fMPS respectively. As explained in the previous section, odd algebra fMPS have the physical property of Majorana edge modes on a chain with open boundary conditions.

\subsection{$\mathds{Z}_2$ group structure}\label{Z2group}

Taking the graded tensor product of two fMPS with odd fermion parity under periodic boundary conditions obviously gives an fMPS with even fermion parity. In the previous section we related the global fermion parity of the fMPS to the type of simple $\mathbb{Z}_2$ graded algebra spanned by the matrices $A^i$. In this section we calculate how the type of simple graded algebra changes under the graded tensor product of two odd fMPS. We start by taking the graded tensor product of two fMPS tensors, and applying the fermionic reordering to obtain the tensor of the new fMPS:
\begin{eqnarray*}
\mathcal{F}\left(\left(\sum_{i\alpha\beta}A^i_{\alpha\beta}|\alpha)|i\rangle (\beta| \right)\otimes_{\mathfrak{g}}\left(\sum_{j\gamma\delta}A'^j_{\gamma\delta}|\gamma)|j\rangle (\delta| \right)  \right)\\
=  \sum_{ij\alpha\beta\gamma\delta}A^i_{\alpha\beta}A'^j_{\gamma\delta}\mathcal{F}\left(|\alpha)|i\rangle (\beta|\otimes_{\mathfrak{g}} |\gamma)|j\rangle (\delta|\right) \\
 = \sum_{ij\alpha\beta\gamma\delta}A^i_{\alpha\beta}A'^j_{\gamma\delta}(-1)^{|\gamma||i|} |\alpha)|\gamma)|i\rangle |j\rangle (\delta|(\beta|
\end{eqnarray*}
From this we see that the tensor components of the fMPS describing the graded tensor product of the individual fMPS are
\begin{equation}
A^{ij}_{(\alpha\gamma)(\beta\delta)} = A^i_{\alpha\beta}A'^j_{\gamma\delta}(-1)^{|\gamma||i|}
\end{equation}
As the tensor product of two simple matrix algebras over $\mathbb{C}$ again forms a simple algebra, it is sufficient to start from the matrices $A^i=y^{|i|}$, i.e.\ those of the Kitaev chain \eqref{Kitaevchain}. We also use a relabeling (permutation) for the virtual bases of the new fMPS tensors in which the even parity states come before the odd parity states, i.e.\ so that the parity matrix $\mathcal{P}$ takes the standard form. We thus obtain:
\begin{align*}
A^{00} = &  \mathds{1}\otimes \mathds{1},&  A^{11}  =&  \mathds{1}\otimes y \\
A^{01} = & y \otimes z,& A^{10}  =&   y \otimes x\, ,
\end{align*}
with $y = \left(\begin{matrix} 0 & 1 \\ -1 & 0\end{matrix}\right), z = \left(\begin{matrix} 1 & 0 \\ 0& -1\end{matrix}\right)$ and $x = \left(\begin{matrix} 0 & 1 \\ 1 & 0\end{matrix}\right)$. The Kitaev chain fMPS with periodic boundary conditions also contains an additional matrix $y$ in the trace defining the coefficients. The reordered graded tensor product $\mathcal{F}\left(y\otimes_{\mathfrak{g}} y \right)$ expressed in the new basis given is by $-z\otimes y$. As the total fermion parity of the tensor product of two odd chains is even, the final fermion contraction induces an extra factor $\mathcal{P}=z\otimes \mathds{1}$. We thus see that the coefficients of the new fMPS are of the form
\begin{align}
\text{tr}\left((\mathds{1}\otimes y)A^{i_1j_1}A^{i_2j_2}\dots A^{i_Nj_N} \right)  \nonumber \\
\equiv -\text{tr}\left(i\Lambda A^{i_1j_1}A^{i_2j_2}\dots A^{i_Nj_N} \right)\, .
\end{align}
The matrices $A^{ij}$ are of the form $y^{|i|+|j|} \otimes B^{i,j}$ where the even subalgebra of the $B$ matrices is spanned by $\mathds{1}$ and $y$ and is thus reducible. This implies the existence of a parity preserving projector $Q = (1+i y)/2 \oplus (1-i y)/2$ and its complement $Q^\perp = (1-i y)/2 \oplus (1+i y)/2$. The non-trivial closure $i\Lambda = -\mathds{1}\otimes y$ makes this state nonzero. The following gauge transform makes this form explicit
\begin{displaymath}
\frac{1}{\sqrt{2}}\begin{bmatrix}
0 & -i & 0 & i\\
0 & 1 & 0 & 1\\
i & 0 & -i & 0\\
1 & 0 & 1 & 0
\end{bmatrix}
\end{displaymath}
and transforms the matrices into
\begin{align*}
A^{00} = &  \mathds{1}\oplus \mathds{1},&  A^{11}  =& (-iz) \oplus (-iz), \\
A^{01} = & (-ix) \oplus (ix) ,& A^{10}  =&  (y) \oplus (-y),\\
\mathcal{P} &= z \oplus z,&i\Lambda = & (iz) \oplus (iz) = i\mathcal{P}\, .
\end{align*}
Although the two individual fMPS tensors in the direct sum decomposition have a different sign in case of odd $|i|$, this sign is irrelevant as the odd matrices appear an even number of times. Hence, up to a global factor $-2i$, the tensor product of two Kitaev chain fMPS takes the standard form of an even fMPS [Eq.~\eqref{evenalgebra}] with matrices
\begin{align*}
A^{00} = &  \mathds{1},&  A^{11}  =& (-iz), \\
A^{01} = & -ix ,& A^{10}  =&  y.
\end{align*}
These matrices clearly span an even simple $\mathbb{Z}_2$ graded algebra. However, we notice that under time reversal for spinless fermions the matrices of the new fMPS transform as 
\begin{align}
\bar{A}^{ij} &= y^TA^{ij}y\\
-i\mathcal{P} &= y^T\left(i\mathcal{P}\right) y\, ,
\end{align}
which implies that there will be Kramers pairs at the ends of an open chain because the global action of time reversal on the chain gets intertwined to an action of $y$ on the virtual indices at the ends and $y^2 = -\mathds{1}$ \cite{Fidkowski3,Turner}. We come back to this point in more detail in section \ref{sec:timereversal}.

\section{Parent Hamiltonian and ground state uniqueness}\label{sec:parent}
In section \ref{bosonicMPS} we explained how every bosonic MPS has a parent Hamiltonian associated to it. We show that this construction carries over directly to the fMPS framework and that the resulting parent Hamiltonian, both for even and odd algebra fMPS, has a unique ground state on a closed chain with periodic boundary conditions. The parent Hamiltonian construction is most cleanly expressed using fermionic tensors, such that the resulting Hamiltonian terms are fermionic operators by construction. The reason for this is that the framework of fermionic tensor networks was set up in such a way (the use of even tensors and at most one odd tensor) that applying the fermionic reordering isomorphism $\mathcal{F}$ before contracting the tensors, as well as the order in which the various contractions are evaluated, has no effect on the outcome. This is precisely what warrants the validity of the popular graphical notation used for bosonic tensor networks, and is thus still valid in the current fermionic context. Even at the level of a single tensor we can apply $\mathcal{F}$ in order to choose a different internal ordering with respect to which the tensor coefficients are defined. This information is not encoded in the graphical notation, but does again not influence the end result, if the tensor coefficients are correctly transformed when going from one particular ordering to another. As such, boxes in the graphical tensor network notation do not denote a single tensor $\mathsf{A}$, but the whole equivalence class $[\mathsf{A}]$ of tensors related by the reordering isomorphism $\mathcal{F}$.

We first construct the parent Hamiltonian for the case of an irreducible even algebra fMPS and start by blocking the physical sites such that the tensors become injective at the single site level, i.e.\ $A_{\alpha,\beta}^i$ as a map from the $D^2$-dimensional space corresponding to $\alpha$ and $\beta$ to the $d$-dimensional space of $i$ is injective. If $(A^+)^i_{\alpha,\beta}$ denotes the Moore-Penrose pseudo-inverse of $A^i_{\alpha,\beta}$, interpreted as matrix with rows $i$ and columns $\alpha,\beta$, then it actually is a left inverse, i.e.
\begin{equation}
\sum_{i} (A^+)^i_{\alpha,\beta} A^i_{\alpha',\beta'} = \delta_{\alpha,\alpha'}\delta_{\beta,\beta'}
\end{equation}
Generically, the left inverse is not unique, but the Moore-Penrose pseudo-inverse is singled out by the condition that, acting on the right, it gives rise to a hermitian projector
\begin{equation}
\sum_{\alpha,\beta}  A^i_{\alpha,\beta} (A^+)^{i'}_{\alpha,\beta}= (P_1)_{i,i'}
\end{equation}
We now lift this definition to a fermionic tensor $\mathsf{A}^+$ as
\begin{equation}
\mathsf{A}^+ = \sum_{i,\alpha,\beta} (-1)^{|\alpha|} (A^+)^i_{\alpha,\beta} |\beta)\otimesg \langle i| \otimesg (\alpha|. \label{eq:fpseudoinv}
\end{equation}
\begin{widetext}
The reason for the additional sign $(-1)^{|\alpha|}$ becomes clear when we contract $\mathsf{A}^+\otimesg \mathsf{A}$ over the physical index and use a fermionic reordering to obtain
\begin{align*}
\mathcal{C}(\mathsf{A}^+\otimesg \mathsf{A})&=\mathcal{C}\big(\sum_{i,\alpha,\alpha',\beta,\beta'} (-1)^{|\alpha|} (A^+)^i_{\alpha,\beta} A^i_{\alpha',\beta'} |\beta)\otimesg \langle i| \otimesg (\alpha| \otimesg |\alpha') \otimesg |i\rangle \otimesg (\beta'|\big)\\
&\stackrel{\mathcal{F}}{\to}\mathcal{C}\big(\sum_{i,\alpha,\alpha',\beta,\beta'} (A^+)^i_{\alpha,\beta} A^i_{\alpha',\beta'} |\alpha')\otimesg (\alpha| \otimesg |\beta)  \otimesg \langle i| \otimesg |i\rangle \otimesg (\beta'|\big)\\
&= (\sum_{\alpha} |\alpha)(\alpha|)\otimesg (\sum_{\beta} |\beta) (\beta|) = \mathds{1}.
\end{align*}
\end{widetext}
Hence, $\mathsf{A}^+$ is a left inverse of $\mathsf{A}$, i.e.\ in any tensor network diagram, we can cancel $\mathsf{A}^+$ and $\mathsf{A}$ when contracted along the physical index.

\begin{figure}
  \centering
    \includegraphics[width=0.48\textwidth]{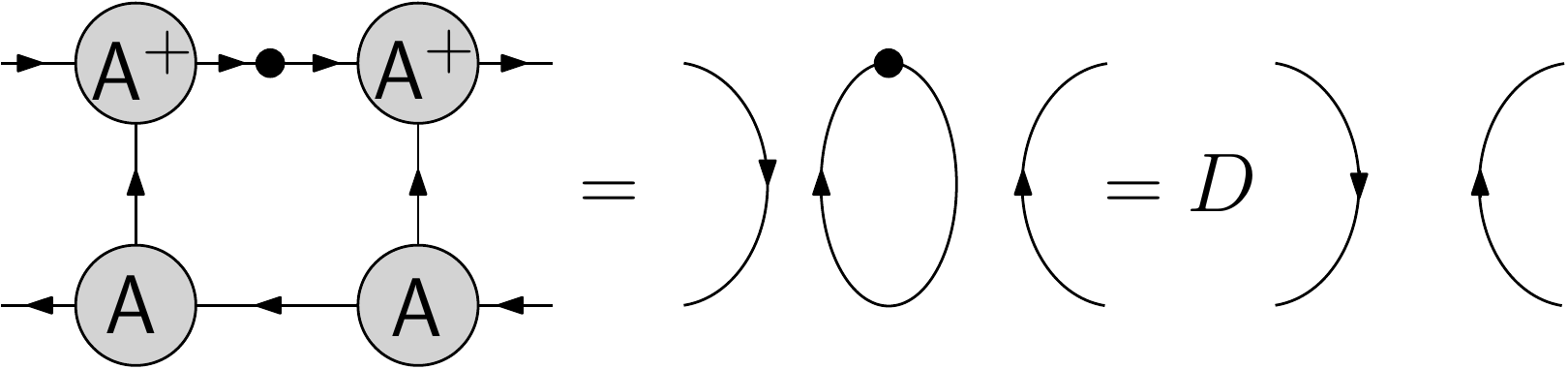}
  \caption{Stability of the injectivity property under contraction of tensors. The figure represents $\mathcal{C}(\mathsf{A}_2^{(-1)}\otimesg \mathsf{A}_2) =D \mathds{1}\otimesg \mathds{1}$, with $D$ the bond dimension. The black dot represents the matrix $\sum_\alpha (-1)^{\alpha}|\alpha)(\alpha|$. The arrows in the diagrammatic notation denote which indices correspond to bra's and which indices correspond to kets in the corresponding super vector spaces.}\label{fig:stability}
\end{figure}

However, upon applying several neighbouring $\mathsf{A}^+$ to the MPS, we cannot simply contract the inner virtual degrees of freedom, as we will get supertraces (as discussed in Section~\ref{sec:super}) which are not merely traces of the identity. This is resolved by simply inserting additional parity uperators when concatenating $\mathsf{A}^+$ tensors. Consider hereto the MPS tensor $\mathsf{A}_2$ of a two-site block, which is defined by contracting a single bond between two MPS tensors $\mathsf{A}\otimesg \mathsf{A}$
\begin{equation}
\mathsf{A}_2 = \mathcal{C}(\mathsf{A}\otimesg \mathsf{A}) = \sum_{\alpha,\beta,\gamma,i,j} A^i_{\alpha,\beta}A^{j}_{\beta,\gamma} |\alpha)|i\rangle |j\rangle (\gamma|.
\end{equation}
By constructing a left inverse $\mathsf{A}_2^{(-1)}$ of $\mathsf{A}_2$ as
\begin{align*}
\mathsf{A}_2^{(-1)} &= \mathcal{C}\big(\mathsf{A}^+ \otimesg (\sum_\beta (-1)^{|\beta|}|\beta)(\beta|) \otimesg \mathsf{A}^+ \big)\\
&= \sum_{i,j,\alpha,\beta,\gamma } (-1)^{|\alpha|} (A^+)^i_{\alpha,\beta}(A^+)^j_{\beta,\gamma} \\ 
& \hspace{19 mm} |\gamma)\otimesg \langle j| \otimesg \langle i| \otimesg (\alpha|
\end{align*}
one can show stability of the injectivity property under contraction of tensors. This is illustrated in figure \ref{fig:stability} using a diagrammatic notation.

A different left inverse
\begin{equation}
\mathsf{A}_2^{+} = \sum_{i,j,\alpha,\beta,\gamma } (-1)^{|\alpha|} (A_2^+)^{i,j}_{\alpha,\gamma} |\gamma)\otimesg \langle j| \otimesg \langle i| \otimesg (\alpha|.
\end{equation}
can be constructed from the pseudo-inverse $(A_2^+)^{i,j}_{\alpha,\gamma}$ of the coefficients $(A_2)^{i,j}_{\alpha,\gamma}=\sum_{\beta}  A^i_{\alpha,\beta}A^{j}_{\beta,\gamma}$, interpreted as matrix with rows $(i,j)$ and columns $(\alpha,\beta)$. While $\mathsf{A}_2^{(-1)}$ can be used to prove the ground state uniqueness (see below), we need to use $\mathsf{A}_2^{+}$ in the construction of the parent Hamiltonian, in order to obtain a hermitian operator. By contracting $\mathsf{A}_2\otimesg \mathsf{A}_2^+$ along the virtual bonds $\alpha$ and $\gamma$, we obtain the orthogonal projector $\mathsf{P}_2$ onto the physical support on two sites as
\begin{eqnarray}
\mathsf{P}_2 & = &\mathcal{C}(\mathsf{A}_2\otimesg \mathsf{A}_2^+) \\
 & = &\sum_{i,i',j,j',\alpha,\gamma} (A_2)^{i'j'}_{\alpha,\gamma}(A_2^+)^{ij}_{\alpha,\gamma} \ket{i'}\ket{j'}\bra{j}\bra{i}.	\nonumber
\end{eqnarray}
The extra factor $(-1)^{|\alpha|}$ was cancelled by the factor picked up from reordering the $(\alpha|$ before contracting it with $|\alpha)$. The resulting fermion operator $\mathsf{P}_2$ is a hermitian projector, as can be verified from the properties of the pseudo-inverse $(A_2^+)^{i,j}_{\alpha,\gamma}$. The parent Hamiltonian is then simply
\begin{equation}\label{parentHam}
H = \sum_{i = 1}^N (\mathds{1} - \mathsf{P}_2)_i\, ,
\end{equation}
where $(1-\mathsf{P}_2)_i$ acts on two consecutive sites $i$ and $i+1 \mod N$. It clearly annihilates the fMPS. Vice versa, when showing uniqueness, we use that the ground subspace of the local terms is of the form $\mathcal{C}(\mathsf{A} \otimesg \mathsf{A}\otimes \mathsf{X})$ with $\mathsf{X}$ an arbitrary tensor with well-defined parity (we try even and odd seperately) on the other sites. That the intersection of all these local ground states only contains the translation invariant even parity fMPS containing only $\mathsf{A}$ tensors follows from proving the intersection and closure property from Ref.~\cite{Schuch}. These proofs can be completely expressed in terms of tensor network diagrams of the type shown in figure \ref{fig:stability} (using the left inverse $\mathsf{A}_2^{(-1)}$) and are therefore equally valid for the bosonic and fermionic case. 

For the odd algebra $A^i=y^{|i|} \otimes B^i$, we first block sites to the point where $B^i_{a,b}$ with $|i|=0$ provides an injective mapping from the $(D/2)^2$ dimensional space labeled by $a$ and $b$ to the $d_e$-dimensional space of $|i|=0$, and separately for $B^i_{a,b}$ with $|i|=1$ \footnote{Note that under blocking two sites n times, we have $d_e^{(n)} = (d_e^{(n-1)})^2+(d_o^{(n-1)})^2$ and $d_o^{(n)}=2d_e^{(n-1)}d_o^{(n-1)}$, so that the ratio $x^{(n)} = d_e^{(n)}/d_o^{(n)}= x^{(n-1)}/2 + 1/2x^{(n-1)}$ converges to $1$ and thus $d_{e,o}^{(n)}\to d^{(n)}/2 = d^{2^n}/2$.}. We here used the notation $\alpha = (|\alpha|,a)$ where $a=1,\ldots D/2$ spans the space in which the matrices $B$ live and does not carry any information of the parity. Let us denote $B^+$ as the pseudo-inverse of $B$ in the two parity sectors individually, i.e.\ $\sum_{|i|=0} (B^+)^i_{a,b} B^i_{a',b'} =\delta_{a,a'} \delta_{b,b'}$ and similarly for the sum over $|i|=1$, whereas $\sum_{a,b} B^i_{a,b} (B^+)^{i'}_{a,b}$ is a hermitian projector in the block corresponding to $|i|=|i'|=0$, as well as when restricted to the block $|i|=|i'|=1$. The Moore-Read pseudo-inverse of $A$ can than be verified to be given by $(A^+)^i_{\alpha,\beta} = \frac{1}{2} (y^{|i|})_{|\alpha|,|\beta|} (B^+)^i_{a,b}$. It indeed satisfies
\begin{align*}
\sum_{\alpha,\beta} A^i_{\alpha,\beta} (A^+)^{i'}_{\alpha,\beta} &= \sum_{\alpha,\beta}\frac{1}{2} (y^{|i|})_{|\alpha|,|\beta|}(y^{|i'|})_{|\alpha|,|\beta|} B^i_{a,b} (B^+)^{i'}_{a,b} \\
&= \delta_{|i|,|i'|} \sum_{a,b} B^i_{a,b} (B^+)^{i'}_{a,b}\\
&= (P_1)_{i,i'}
\end{align*}
with $P_1$ a hermitian projector. Note that the restriction to $|i|=|i'|$ follows from the $y$ part in $A$ and $A^+$. Acting on the left, by evaluating the sum over $|i|=0$ and $|i|=1$ separately, it results in
\begin{align*}
\sum_{i} &(A^+)^{i}_{\alpha',\beta'}A^i_{\alpha,\beta}  \\
&= \frac{1}{2}[\delta_{|\alpha|,|\beta|}\delta_{|\alpha'|,|\beta'|} + y_{|\alpha|,|\beta|} y_{|\alpha'|,|\beta'|}] \delta_{a,a'}\delta_{b,b'}\\
&= \frac{1}{2}[\delta_{|\alpha|,|\alpha'|}\delta_{|\beta|,|\beta'|} + y_{|\alpha|,|\alpha'|} y_{|\beta|,|\beta'|}] \delta_{a,a'}\delta_{b,b'}\\
\end{align*}
\begin{widetext}
Again defining the fermionic tensor $\mathsf{A}^+$ using Eq.~\eqref{eq:fpseudoinv}, we obtain
\begin{align*}
\mathcal{C}(\mathsf{A}^+\otimesg \mathsf{A}) \stackrel{\mathcal{F}}{\to}& \sum_{\alpha,\beta,\alpha',\beta'} \frac{1}{2}[\delta_{|\alpha|,|\alpha'|}\delta_{|\beta|,|\beta'|} + y_{|\alpha|,|\alpha'|} y_{|\beta|,|\beta'|}] \delta_{a,a'}\delta_{b,b'} |\alpha')(\alpha| \otimesg |\beta)(\beta'|\\
&= \frac{1}{2}\left[\mathds{1} \otimesg \mathds{1} - \mathsf{Y} \otimesg \mathsf{Y}\right]\, ,
\end{align*}
where the minus sign in the second term originates from $y=-y^{T}$.
\end{widetext}
Hence, $\mathsf{A}^+$ acts as a left pseudo-inverse onto the subspace where the tensor is injective. As in the even algebra case, we obtain a left pseudo-inverse $\mathsf{A}_2^{(-1)}$ on two sites by combining $\mathsf{A}^+\otimesg \mathsf{A}^+$ with a parity operator on the internal virtual bond. The latter gives rise to a normal trace, so that we obtain
\begin{align*}
\mathcal{C}(\mathsf{A}_2^{-1}\otimesg \mathsf{A}_2) \sim & \mathds{1} \otimesg \tr(\mathds{1}) \otimesg \mathds{1} - \mathsf{Y} \otimesg \tr(\mathsf{Y}) \otimesg \mathds{1} \\
&- \mathds{1} \otimesg \tr(\mathsf{Y}) \otimesg \mathsf{Y} + \mathsf{Y} \otimesg \tr(\mathsf{Y}\mathsf{Y}) \otimesg \mathsf{Y}\\
\sim& \mathds{1} \otimesg \mathds{1}-\mathsf{Y} \otimesg \mathsf{Y},
\end{align*}
i.e.\ the left inverse property is stable under blocking. The parent Hamiltonian is constructed in the same way as in the even case using the pseudo-inverse $\mathsf{A}_{2}^+$. Let us make the explicit exercise for the fermionic MPS representing the Kitaev chain, with tensor $A^i = y^{|i|}$ and thus $B^0 = B^1 = 1$. On a single site, this tensor already satisfies the injectivity condition, so no extra blocking is needed. We obtain $B_2^{00} = B_2^{01} = B_2^{10} = - B_2^{11} = 1$. The pseudo-inverse $B_2^{+}$ is constructed using the parts of even and odd $|i|+|j|$ separately, so we obtain $(B_2^+)^{00}=- (B_2^+)^{11} = 1/2$ for the even part and $(B_2^+)^{01} = (B_2^+)^{10} = 1/2$ for the odd part. We then find $(A_2^+)^{00} = - (A_2^+)^{11} = \mathds{1}/4$ and $(A_2^+)^{01} = (A_2^+)^{10} = y/4$. The projector onto the physical support on two sites is then given by
\begin{equation*}
\mathsf{P}_2 = \frac{1}{2} (|00\rangle - |11\rangle)(\langle00|-\langle 11|) + \frac{1}{2} (|01\rangle + |10\rangle)(\langle 01|+\langle 10|) 
\end{equation*}
and by rewriting it in terms of creation and annihilation operators, we obtain the projector $P_j$ defined for the Kitaev Hamiltonian in Eq. \eqref{kitaevprojector}, up to a constant.

To prove uniqueness of the ground state, we recycle the intersection and closure property of $\mathbb{Z}_2$-injective MPS, as discussed in Ref.~\cite{Schuch}. These are proven diagrammatically and are thus equally valid in the fermionic case. The only difference is in the final conclusion. We find again that the most general fermionic tensor that can be used to close the fMPS with periodic boundary conditions is a two-dimensional linear combination, namely of $\mathds{1}$ and $\mathsf{Y}$. However, when investigating the linear independence of these two closures, the first one turns out to give rise to a zero state, as already discussed in the previous section. So we end up with a unique ground state for the parent Hamiltonian~\eqref{parentHam} corresponding to the closure $\mathsf{Y}$, and thus to a state with odd fermion parity.

Let us now discuss the odd algebra fMPS case in more physical terms. To recapitulate, after evaluating the contraction of the virtual bonds, the odd algebra fMPS takes the form
\begin{equation}
|\psi\rangle = \sum_{\{i_j\}}\text{tr}\left(YA^{i_1}A^{i_2}\dots A^{i_N} \right)|i_1\rangle|i_2\rangle\dots |i_N\rangle\, ,
\end{equation}
with $Y = y\otimes \mathds{1}$ and $A^i = y^{|i|}\otimes B^i$. In the bosonic case, the state
\begin{equation}
|\varphi\rangle = \sum_{\{i_j\}}\text{tr}\left(A^{i_1}A^{i_2}\dots A^{i_N} \right)|i_1\rangle|i_2\rangle\dots |i_N\rangle\, .
\end{equation}
would be another ground state. However, if we close the fMPS with an fermionic identity $\sum_{\alpha} |\alpha)_N(\alpha|_1$ and evaluate the virtual contractions, we pick up an additional matrix $\mathcal{P}$ which renders the state zero, as discussed above. Vice versa, we can try to obtain the state $\ket{\varphi}$ as a fermionic state with even fermion parity. This requires that we start from a fermionic tensor network that is closed with an extra $\sum_{\alpha} (-1)^{|\alpha|} | \alpha)_N (\alpha|_1$ factor. However, this factor is not in the center and can therefore not be moved to other positions. Indeed, an even fermionic state $\ket{\varphi}$ is not translation invariant (see appendix \ref{app:translation}) and the position of this factor will be detected by the parent Hamiltonian and cost energy. Hence, $|\varphi\rangle$ is not a ground state of the parent Hamiltonian. We note, however, that $|\varphi\rangle$ is invariant under $T_{AP}$ (see appendix \ref{app:translation} for the definition of $T_{AP}$). Hence, $\ket{\varphi}$ would be the ground state of the Hamiltonian obtained after inserting a $\pi$-flux. Indeed, Majorana chains on a ring change their ground state parity under insertion of a $\pi$-flux, which is the characterizing topological bulk response.

\section{Entanglement spectrum and Majorana modes}\label{sec:entanglement}

It was shown in Ref.~\cite{Fidkowski4} that within the mean-field BCS description of superconductors the presence of Majorana zero modes leads to a two-fold degeneracy in the entanglement spectrum of the ground state wave function. We will use the fMPS description of Majorana chains to show why this remains true beyond the mean-field approximation. We note that the two-fold degeneracy in the entanglement spectrum of interacting Majorana chains was also discussed in Ref. \cite{Turner}. Let us first define the transfer matrix
\begin{equation}\label{transfermatrix}
\mathbb{E}_{(\alpha\gamma)(\beta\delta)} = \sum_i A^i_{\alpha\beta}\bar{A}^i_{\gamma\delta}\, ,
\end{equation}
which is a $D^2\times D^2$ matrix. It is an important object since it appears in every calculation of expectation values with (f)MPS. Normalization of the (f)MPS implies that the largest eigenvalue of $\mathbb{E}$ has norm one. If the $A^i$ span a simple $D\times D$ matrix algebra, one can show that this largest eigenvalue is unique and that the associated left and right fixed points are positive matrices. Given this fact, one can always perform a gauge transformation such that one of the two fixed points, say the left fixed point, is the identity matrix and the other, right, fixed point a positive diagonal matrix \cite{Fannes,PerezGarcia}. For an odd algebra fMPS the matrices are of the form $A^i=y^{|i|}\otimes B^i$. Let us first consider 
\begin{equation}
\mathbb{E}'_{(\alpha\gamma)(\beta\delta)} = \sum_i B^i_{\alpha\beta}\bar{B}^i_{\gamma\delta}\, .
\end{equation}
If the odd algebra fMPS is irreducible then we can work in the gauge described above such that the $(D/2)^2\times (D/2)^2$ matrix $\mathbb{E}'$ has a unique left fixed point given by the identity and a unique right fixed point given by a diagonal positive matrix $\Lambda$. The fixed points of the full transfer matrix $\mathbb{E}$ will of course not be unique since $A^i = y^{|i|}\otimes B^i$ do not span a simple matrix algebra. As it turns out, the eigenspace corresponding to eigenvalue 1 is two-dimensional. Using $\Lambda$ we can easily find two orthogonal right fixed points of $\mathbb{E}$:
\begin{align}
R_e & = \mathds{1}\otimes \Lambda \\
R_o & = y \otimes \Lambda\, .
\end{align}
Similarly, two orthogonal left fixed points of $\mathbb{E}$ are given by
\begin{align}
L_e & = \mathds{1}\otimes \mathds{1} \\
L_o & = y \otimes \mathds{1}\, .
\end{align}
Let us now consider an odd algebra fMPS defined on a chain of length $N$ with open boundaries
\begin{equation}
|\psi\rangle = \sum_{\{i\}} v_L^TA^{i_1}A^{i_2}\dots A^{i_N}v_R\, |i_1\rangle|i_2\rangle\dots|i_N\rangle\, ,
\end{equation}
where $v_L$ and $v_R$ are two $D$-dimensional vectors used to close the virtual indices on the boundary. The Hermitian conjugate of $|\psi\rangle$ is given by
\begin{equation}
\langle \psi| = \sum_{\{i\}} \bar{v}_L^T\bar{A}^{i_1}\bar{A}^{i_2}\dots \bar{A}^{i_N}\bar{v}_R\, \langle i_N|\dots\langle i_2|\langle i_1|\, .
\end{equation}
We will now divide the chain in two, where the first $N/2$ sites are in subsystem $A$ and the $N/2$ sites on the right constitute subsystem $B$. 
\begin{widetext}
The reduced density matrix of subsystem $A$ is then defined as \footnote{$\rho_A$ is a positive matrix obtained from tracing out the degrees of freedom in region $B$, starting from $|\psi\rangle\langle\psi|$. Note that in this fermionic setting, tracing is not obtained by simply contracting using $\mathcal{C}$, as the latter gives rise to the supertrace as discussed in Section~\ref{sec:super}. Instead, we first have to apply the corresponding parity operator.}
\begin{align}
\rho_A \equiv &  \sum_{i_1\dots i_{N/2},i'_1\dots i'_{N/2}}\Big(\sum_{i_{N/2+1}\dots i_N} ( v_L^TA^{i_1}\dots A^{i_{N/2}}A^{i_{N/2+1}}\dots A^{i_N}v_R) \nonumber\\
& (\bar{v}_L^T\bar{A}^{i'_1}\dots \bar{A}^{i'_{N/2}}\bar{A}^{i_{N/2+1}}\dots \bar{A}^{i_N}\bar{v}_R)\Big)|i_1\rangle\dots|i_{N/2}\rangle\langle i'_{N/2}|\dots \langle i'_1|\, .
\end{align}
In the expression for $\rho_A$ we recognize $N/2$ times the application of $\mathbb{E}$ to $v_R\otimes \bar{v}_R$. Since $v_R$ has a well-defined fermion parity, $v_R\otimes \bar{v}_R$ is even such that for $N$ going to infinity we get
\begin{equation}
\lim_{N\rightarrow \infty} \sum_{\beta\delta}\left(\mathbb{E}^{N/2}\right)_{(\alpha\gamma)(\beta\delta)}(v_R)_\beta(\bar{v}_R)_\delta = \mathds{1}\otimes \Lambda = R_e\, .
\end{equation}
Because we are only interested in this limit we can rewrite the reduced density matrix of subsystem $A$ as
\begin{align}
\rho_A =  \sum_{i_1\dots i_{N/2},i'_1\dots i'_{N/2}}\Big(\sum_{\alpha\beta}( v_L^TA^{i_1}\dots A^{i_{N/2}})_\alpha (R_e)_{\alpha\beta}
(\bar{v}_L^T\bar{A}^{i'_1}\dots \bar{A}^{i'_{N/2}})_\beta\Big) 
|i_1\rangle\dots|i_{N/2}\rangle\langle i'_{N/2}|\dots \langle i'_1|\, .
\end{align}
Using the fact that the spectra of $AB$ and $BA$ for two matrices $A$ and $B$ are the same, we see that the eigenvalues of $\rho_A$ are the same as the eigenvalues of the matrix
\begin{equation}
E_{\alpha\beta} = \sum_\gamma \left(\sum_{i_1\dots i_{N/2}}( v_L^TA^{i_1}\dots A^{i_{N/2}})_\gamma (\bar{v}_L^T\bar{A}^{i_1}\dots \bar{A}^{i_{N/2}})_\alpha \right)(R_e)_{\gamma\beta}\, ,
\end{equation}
where we again recognize multiple applications of the transfer matrix. 
\end{widetext}
Using
\begin{equation}
\lim_{N\rightarrow \infty} \sum_{\kappa\lambda}(v_L^T)_\kappa (\bar{v}_L^T)_\lambda \left(\mathbb{E}^{N/2}\right)_{(\kappa\lambda)(\gamma\alpha)} = (\mathds{1}\otimes \mathds{1})_{\gamma\alpha} = (L_e)_{\gamma\alpha}
\end{equation}
we find that in the large $N$ limit
\begin{equation}
\text{spec}(\rho_A) = \text{spec}(L_e^TR_e)\, .
\end{equation}
Since $L_e^TR_e = \mathds{1}\otimes\Lambda$, we indeed see that the entanglement spectrum is two-fold degenerate.

\section{Time reversal symmetry and the $\mathds{Z}_8$ classification}\label{sec:timereversal}

In section \ref{Z2group} we saw that the graded tensor product of two Kitaev chains is non-trivial when time reversal is included because of the appearance of Kramers pairs at the ends of open chains. To recapitulate, we showed that the tensors describing the graded tensor product of the ground states of two Kitaev chains satisfy
\begin{equation}
\bar{A}^{ij} = \mathcal{T}^{-1}A^{ij}\mathcal{T}\, ,
\end{equation} 
with $\mathcal{T} = y$ and $\mathcal{P} = z$. This relation will also hold for the tensors describing the ground state of $n$ Kitaev chains, but with different expressions for $\mathcal{T}$. Just as in the MPS description of symmetry-protected phases in spin chains, we can extract discrete invariants from $\mathcal{T}$. For an even algebra fMPS these invariants are $\kappa \in \{0,1\}$ and $\mu \in \{0,1\}$, defined via
\begin{align}
\mathcal{T}\bar{\mathcal{T}} &= (-1)^\kappa \mathds{1}\\
\mathcal{T}\mathcal{P} &= (-1)^{\mu}\mathcal{P}\mathcal{T}\, .
\end{align}
The first requirement stems from the fact that time reversal for spinless fermions squares to the identity. Because time reversal is anti-unitary $(-1)^\kappa$ cannot be absorbed in a redefinition of $\mathcal{T}$. The invariant $\mu$ denotes whether $\mathcal{T}$ has even or odd fermion parity. So for $n=2$ we see that $\kappa = \mu = 1$. 

For an odd algebra fMPS we have to define different invariants. Let us for example consider the case $n=1$, i.e. a single Kitaev chain described by the matrices $A^0=\mathds{1}$ and $A^1 = y$. We easily see that
\begin{equation}
\bar{A}^i = A^i\, ,
\end{equation}
so $\mathcal{T} = \mathds{1}$. However, for an odd algebra fMPS the graded algebra spanned by the $A^i$ has a graded center consisting of multiples of $\mathds{1}$ and $Y (=y)$. So we could equivalently choose $\mathcal{T} = y$. Since $y^2 = -\mathds{1}$ this ambiguity makes the invariants as defined for an even algebra fMPS ill-defined for an odd algebra fMPS. However, we can use this ambiguity to always make $\mathcal{T}$ even, i.e. of the two choices for $\mathcal{T}$ given by $\mathcal{T}_e$ and $\mathcal{T}_o = Y\mathcal{T}_e$ we always pick $\mathcal{T}_e$ and then define $\kappa$ as
\begin{equation}
\mathcal{T}_e\bar{\mathcal{T}}_e = (-1)^\kappa \mathds{1}\, .
\end{equation}
To define the $\mu$ invariant we note that for the odd algebra fMPS with periodic boundary conditions to be invariant under time reversal it should hold that:
\begin{equation}\label{mudefinition}
\mathcal{T}\bar{Y} = (-1)^\mu Y\mathcal{T}\, .
\end{equation}
Note that we could change $\mu$ by redefining $Y\rightarrow iY$. For this reason equation \eqref{mudefinition} should be considered together with the requirement $Y^2 = -\mathds{1}$ for $\mu$ to be a true invariant. So we see that $\mu = \kappa = 0$ for $n=1$. The invariants obtained for the even and odd algebra fMPS give rise eight different possibilities, implying that there are eight different spinless superconductors under the protection of time reversal. In the remainder of this section we will study these invariants for the tensors describing the graded tensor product of $n$ copies of the Kitaev chain ground state. In this way we obtain the group structure of time reversal symmetric fMPS under stacking. Let us start with $n=3$.

\emph{\underline{n = 3}}

The matrices of the fMPS ground state of 3 copies of the Kitaev chain are
\begin{align}
\begin{matrix}
A = \mathds{1}\otimes\mathds{1} &B = \mathds{1}\otimes y & C = -iz\otimes x & D = -iz \otimes z \\
E = y\otimes x & F =  y \otimes z & G = -ix \otimes \mathds{1} & H =-ix \otimes y
\end{matrix}
\end{align}
The extra matrix in the trace expression for the fMPS coefficients \eqref{oddfMPS} becomes
\begin{equation}
\mathcal{F}(y \otimes_{\mathfrak{g}} i\mathds{1}) = ix \otimes y = - H\, .
\end{equation}
As expected, these matrices span an odd graded algebra because its graded center contains the odd element $H$. However, these matrices do not take on the standard form $y^{|i|}\otimes A^i$ for an odd graded algebra. To make the odd algebra structure explicit we note that the fMPS coefficients are fully determined by the following properties of the traceless matrices:
\begin{align}
A^2 & = H^2 = \mathds{1}\\ 
B^2 & = C^2 = D^2 = E^2 = F^2 = G^2 = -\mathds{1}\\
[B,G] & = [C,F] = [D,E] = 0\\
[A,X] & = [H,X] = 0\;\;\;\; \forall X \in \{A,B,\dots ,H\}\, ,
\end{align}
and all other elements anti-commuting with each other. We can use a different representation of this algebra by choosing the matrices
\begin{align}
\begin{matrix}
A = \mathds{1}\otimes\mathds{1} &B = \mathds{1}\otimes y & C = \mathds{1}\otimes iz & D =\mathds{1}\otimes ix \\
E = y\otimes x & F =  y \otimes z & G = y \otimes iy & H = y \otimes i \mathds{1}\, ,
\end{matrix}
\end{align}
which are of the standard form $y^{|i|}\otimes A^i$. Indeed, these representations are related by the following gauge transformation
\begin{displaymath}
G = \begin{bmatrix}
\mathds{1}&0\\0&-y 
\end{bmatrix} \times \left(\mathds{1} \otimes \frac{1}{\sqrt{2}} \begin{bmatrix}1 & 1\\-1 & 1\end{bmatrix}\right).
\end{displaymath}
We will not do this step for the other cases below since it is not required to find the invariants associated to time reversal. One can now easily check that the even virtual time reversal action is of the form
\begin{equation}
\mathcal{T}_e = z \otimes y\, ,
\end{equation}
implying that for $n=3$
\begin{equation}
\mathcal{T}_e\bar{\mathcal{T}}_e = -\mathds{1} \;\;\;\;\text{ and }\;\;\;\; \mathcal{T}\bar{Y} = -Y\mathcal{T}\, ,
\end{equation}
with $Y = y \otimes \mathds{1}$.

\emph{\underline{n = 4}}

We can obtain the fMPS tensor for the ground state of four copies of the Kitaev chain by taking the graded tensor product of the tensors corresponding to the $n=2$ case. In this way we find the matrices
\begin{equation}
\begin{matrix}
 \mathds{1}\otimes \mathds{1}  & -i\mathds{1}\otimes z & -iy\otimes y & y \otimes x \\
-i z\otimes z & -z\otimes\mathds{1} &  -x\otimes x & -ix\otimes y \\
-ix \otimes \mathds{1} & -x\otimes z & -z\otimes y & -i z\otimes x \\
y\otimes z & -iy\otimes\mathds{1}  & -i\mathds{1}\otimes x & \mathds{1}\otimes y
\end{matrix}
\end{equation}
These span an even simple graded algebra and the virtual time reversal action is
\begin{equation}
\mathcal{T} = z\otimes y\, .
\end{equation}
So the invariants are
\begin{equation}
\mathcal{T}\bar{\mathcal{T}}= -\mathds{1}\;\;\;\;\text{ and }\;\;\; \mathcal{T}\mathcal{P} = \mathcal{P}\mathcal{T} 
\end{equation}

\emph{\underline{n = 5}}

For $n=5$ we combine the tensors of the $n=2$ and $n=3$ case and find following matrices
\begin{equation}
\begin{matrix}
 \mathds{1}\otimes \mathds{1}\otimes\mathds{1}  &-iy\otimes y\otimes \mathds{1} & -i\mathds{1}\otimes z \otimes \mathds{1} & y\otimes x \otimes \mathds{1} \\
-i z\otimes z\otimes x & -x\otimes x \otimes x &  -z\otimes\mathds{1}\otimes x & -ix\otimes y \otimes x \\
y\otimes z \otimes x & -i\mathds{1}\otimes x\otimes x & -iy\otimes\mathds{1}\otimes x & \mathds{1}\otimes y \otimes x \\
-ix\otimes\mathds{1}\otimes\mathds{1}  & -z\otimes y \otimes \mathds{1} & -x\otimes z \otimes \mathds{1} & -iz \otimes x \otimes \mathds{1}\, ,
\end{matrix}
\end{equation}
and all elements obtained by right multiplication with $\mathds{1}\otimes\mathds{1}\otimes y$. We thus find that the even virtual symmetry action $\mathcal{T}_e$ and the central odd element $Y$ are given by
\begin{align}
\mathcal{T}_e &= z \otimes y \otimes \mathds{1} \\
Y &= x \otimes z \otimes y\, .
\end{align}
So the invariants are given by
\begin{equation}
\mathcal{T}_e\bar{\mathcal{T}}_e = -\mathds{1}\;\;\;\;\text{ and }\;\;\;\; \mathcal{T}\bar{Y} = Y\mathcal{T}\, .
\end{equation}

\emph{\underline{n = 6}}

By combining the matrices of $n = 2$ and $n=4$ we find that the matrices for $n=6$ are generated by following elements
\begin{equation}
\begin{matrix}
-ix\otimes\mathds{1}\otimes\mathds{1} & -iz\otimes z\otimes z\\
-i\mathds{1}\otimes\mathds{1}\otimes z & y\otimes x \otimes \mathds{1}\\
-iy\otimes z \otimes y & i\mathds{1}\otimes z \otimes \mathds{1}
\end{matrix}
\end{equation}
We thus see that the virtual time reversal symmetry action $\mathcal{T}$ is given by
\begin{equation}
\mathcal{T} = y \otimes x \otimes y\, ,
\end{equation}
leading to the invariants
\begin{equation}
\mathcal{T}\bar{\mathcal{T}} = 1\;\;\;\;\text{ and }\;\;\;\; \mathcal{T}\mathcal{P} = -\mathcal{P}\mathcal{T}\, .
\end{equation}

\emph{\underline{n = 7}}

Taking the graded tensor product of $n = 1$ and $n=6$ we find that the matrices of the ground state fMPS of seven copies of the Kitaev chain are generated by 
\begin{equation}
\begin{matrix}
-ix\otimes\mathds{1}\otimes\mathds{1}\otimes\mathds{1} & y\otimes x\otimes\mathds{1}\otimes\mathds{1}\\
-i\mathds{1}\otimes\mathds{1}\otimes\mathds{1}\otimes z & -iy\otimes z\otimes z \otimes y\\
y\otimes z \otimes x \otimes \mathds{1} & i\mathds{1}\otimes\mathds{1}\otimes z \otimes \mathds{1}\\
\mathds{1}\otimes\mathds{1}\otimes y \otimes z & 
\end{matrix}
\end{equation}
The even virtual time reversal action and odd central element corresponding to this odd graded simple algebra are
\begin{align}
\mathcal{T}_e & = z \otimes y \otimes x \otimes y \\
Y &= x\otimes y \otimes \mathds{1}\otimes\mathds{1}\, ,
\end{align}
giving the invariants
\begin{equation}
\mathcal{T}_e\bar{\mathcal{T}}_e = \mathds{1}\;\;\;\;\text{ and }\;\;\;\;  \mathcal{T}\bar{Y} = -Y\mathcal{T}\, .
\end{equation}

\emph{\underline{n = 8}}

For $n=8$ we find by combining $n = 4$ and $n=4$ that the fMPS matrices are generated by
\begin{equation}
\begin{matrix}
iz\otimes z\otimes \mathds{1}\otimes z & i\mathds{1}\otimes z \otimes z \otimes \mathds{1}\\
ix\otimes\mathds{1}\otimes\mathds{1}\otimes\mathds{1} & i y\otimes y \otimes \mathds{1}\otimes\mathds{1}\\
i\mathds{1}\otimes\mathds{1}\otimes\mathds{1}\otimes z & i\mathds{1}\otimes\mathds{1} \otimes z \otimes\mathds{1}\\
iy\otimes z \otimes \mathds{1} \otimes y & iy\otimes x \otimes y \otimes \mathds{1}\, .
\end{matrix}
\end{equation}
We thus see that
\begin{equation}
\mathcal{T} = z \otimes \mathds{1} \otimes y \otimes y\, ,
\end{equation}
such that
\begin{equation}
\mathcal{T}\bar{\mathcal{T}} = 1\;\;\;\;\text{ and }\;\;\;\; \mathcal{T}\mathcal{P} = \mathcal{P}\mathcal{T}\, .
\end{equation}
So for $n=8$ we obtain an even algebra fMPS with $\kappa = \mu =0$. From this we conclude that taking eight copies of the Kitaev chain ground state results in a trivial time reversal symmetric state, implying that the time reversal invariant one-dimensional spinless superconductors form a $\mathbb{Z}_8$ group under stacking \cite{Fidkowski3,Fidkowski2,Turner}.

In section \ref{sec:irreducibility} we mentioned that any odd algebra fMPS can be gauge transformed into the form $A^i = y^{|i|} \otimes B^i$ and is thus trivially time-reversal invariant ($\mathcal{T}_e = \mathds{1}$ and thus $\kappa=0$) if all the matrices $B^i$ are real. The Kitaev fMPS, which has $B^i=1$, was the simplest starting point to generate the $\mathbb{Z}_8$ group of time-reversal invariant fMPS by stacking. However, also if $B^i$ is real for $|i|=0$ and purely imaginary for $|i|=1$ we obtain a state that is trivially time-reversal invariant. This is most easily observed by writing $A^j = (iy)^{|j|} \otimes C^j$ with all the $C^j$ real and performing a unitary gauge transform $u\otimes\mathds{1}$ to a different standard form $A^j = x^{|j|} \otimes C^j$ where all the matrices are again completely real. The simplest case is obtained with $C^j=1$, which would provide an equally simple starting point (i.e.\ fMPS with $D=2$) to build the $\mathbb{Z}_8$ group table. We therefore dub this fMPS the type 2 Majorana chain (and refer to the original Kitaev chain as the type 1 Majorana chain). In this type 2 standard form $A^i = x^{|i|} \otimes C^i$, the odd central element is given by $i x$, where the imaginary factor is included to have $(ix)^2=-1$. As mentioned above, this was required in order to obtain $\mu$ as an invariant. With $\mathcal{T}_e = \mathds{1}$, we thus obtain
\begin{equation}
\mathcal{T}_e\bar{\mathcal{T}}_e = \mathds{1}\;\;\;\;\text{ and }\;\;\;\; \mathcal{T}(\overline{ix}) = -ix\mathcal{T}\, .
\end{equation}
and thus $\kappa=0$ and $\mu=1$. We can now do the same steps as for the type 1 (Kitaev) chain and calculate the invariants for $n'$ copies of the type 2 chain. The result of these calculations, together with the invariants for $n$ copies of the Kitaev chain are presented in Table \ref{TRtable}, where $\epsilon \in \{0,1\}$ denotes whether $n$ copies of the Kitaev chain ground state correspond to an even or odd algebra fMPS.

\begin{table*}
\begin{tabular}{|c | c | c | c | c |}
  \hline			
  \# Type 1 (Kitaev) chains ($A^1 = y$)&  $\epsilon$ & $ \kappa $ & $\mu$ & \# Type 2 chains $(A^1 = x)$ \\ \hline
  $n$ = 1 & 1 & 0 & 0  & $n'$ = 7\\ \hline
  $n$ = 2 & 0 & 1 & 1 & $n'$ = 6\\ \hline
  $n$ = 3 & 1 & 1 & 1 & $n'$ = 5\\ \hline
  $n$ = 4 & 0 & 1 & 0 & $n'$ = 4\\ \hline
  $n$ = 5 & 1 & 1 & 0 & $n'$ = 3\\ \hline
  $n$ = 6 & 0 & 0 & 1 & $n'$ = 2 \\ \hline
  $n$ = 7 & 1 & 0 & 1 & $n'$ = 1\\ \hline
  $n$ = 8 & 0 & 0 & 0 & $n'$ = 8\\
  \hline  
\end{tabular}
\caption{Invariants for $n$ copies of the type 1 or $n'$ copies of the type 2 Majorana chain. $\epsilon$ denotes the simple $\mathbb{Z}_2$ graded algebra type of the fMPS tensor. $\kappa$ and $\mu$ are related to time reversal.}
\label{TRtable}
\end{table*}

The matrices that build up the ground state of $n$ Kitaev chains form a representation of the real Clifford algebra $Cl_{n,0}$. This can easily be seen by explicitly identifying the anti-commuting generators that square to $-\mathds{1}$, as explicitly denoted in Table~\ref{Cliffordtable}. If we consider $n'$ type 2 chains one can similarly check that the ground state matrices correspond to the real Clifford algebra $Cl_{0,n'}$. Using following relations
\begin{align}
& Cl_{p+1,q+1}  \simeq Cl_{p,q}\otimes \mathbb{R}(2) \\
& Cl_{p+8,q} \simeq Cl_{p,q+8} \simeq Cl_{p,q}\otimes \mathbb{R}(16)\, ,
\end{align}
where $\mathbb{R}(n)$ denotes the algebra of real $n$ by $n$ matrices, one can then give an alternative explanation for the mod $8$ periodicity under stacking. It also shows the equivalence of $n=7$ and $n'=1$.

\begin{table*}
\begin{tabular}{|c|c|c|}
\hline
$n=1$ & $y$ & $Cl_{1,0}$ \\ \hline
$n =2$ & $iz$, $ix$ & $Cl_{2,0}$ \\ \hline
$n=3$ & $\mathds{1}\otimes y$, $\mathds{1}\otimes iz$, $y\otimes x$ & $Cl_{3,0}$ \\ \hline
$n=4$ & $i\mathds{1}\otimes z$, $iy\otimes y$, $z\otimes y$, $i\mathds{1}\otimes x$ & $Cl_{4,0}$ \\ \hline
$n=5$ & $i\mathds{1}\otimes z\otimes\mathds{1}$, $y\otimes x\otimes\mathds{1}$, $\mathds{1}\otimes y \otimes x$ & $Cl_{5,0}$ \\ 
 &$iz\otimes x \otimes \mathds{1}$, $\mathds{1}\otimes y \otimes z$ & \\ \hline
$n=6$ & $iy\otimes z \otimes y$, $ix\otimes\mathds{1}\otimes\mathds{1}$, $iz\otimes \mathds{1}\otimes\mathds{1}$ & $Cl_{6,0}$ \\ 
 &$y\otimes x \otimes \mathds{1}$, $y\otimes z \otimes x$, $y\otimes z \otimes z$ & \\
\hline
$n=7$ & $ix\otimes\mathds{1}\otimes\mathds{1}\otimes\mathds{1}$, $y\otimes z \otimes x \otimes \mathds{1}$, $y\otimes x \otimes \mathds{1}\otimes \mathds{1}$,$iy\otimes z \otimes z \otimes y$ & $Cl_{7,0}$ \\ 
 &$iz\otimes x \otimes z \otimes z$, $iz\otimes x \otimes z \otimes z$, $z\otimes x \otimes y\otimes \mathds{1}$ & \\
\hline
$n=8$ & $iy\otimes z \otimes \mathds{1}\otimes y$, $iy\otimes x \otimes y \otimes \mathds{1}$,
$iy\otimes y \otimes \mathds{1}\otimes\mathds{1}$, $ix\otimes\mathds{1}\otimes\mathds{1}\otimes\mathds{1}$ & $Cl_{8,0}$\\
 & $y\otimes z \otimes \mathds{1}\otimes x$, $y\otimes x \otimes x \otimes\mathds{1}$, $y\otimes x\otimes z \otimes \mathds{1}$, $y\otimes z \otimes \mathds{1}\otimes z$ & \\
\hline
\end{tabular}
\caption{Anti-commuting generators that square to $-1$ for $n$ copies of the Kitaev chain, thus providing a representation of the real Clifford algebra $Cl_{n,0}$.}
\label{Cliffordtable}
\end{table*}

\section{General on-site symmetries}\label{sec:onsite}

After having studied time-reversal for spinless fermions in the previous section we will now turn to general on-site symmetries. The goal is to find the invariants and group structure of symmetric phases. For fermionic systems the on-site symmetry group $G$ always contains the central $\mathbb{Z}_2$ subgroup $\{I,P\}$, where $I$ is the trivial group element and $P$ is fermion parity. We denote the on-site symmetry representation as $U(g)$. Note that $U(g)P = PU(g),\forall g\in G$. Acting with the symmetry on the MPS should leave it invariant, possibly up to a phase. The fundamental theorem of MPS \cite{PerezGarcia,PerezGarcia2,Cirac} can also be used in the fermionic case and states that two MPS are the same for every length of the lattice on which they are defined iff their tensors are related by a gauge transformation. Concretely, this implies that the invariance of a MPS under a unitary on-site symmetry is translated into the following local tensor relation:
\begin{equation}\label{symm1}
\sum_{j}U(g)_{ij}A^{j} = e^{i\theta(g)}V(g)^{-1} A^iV(g)\, .
\end{equation}
For an anti-unitary symmetry we get a similar condition:
\begin{equation}\label{symm2}
\sum_{j}U(g)_{ij}\bar{A}^{j} = e^{i\theta(g)}V(g)^{-1} A^iV(g)\, .
\end{equation}
In appendix \ref{app:parity} we show that without loss of generality one can assume that $V(g)$ has a well-defined parity, i.e. $\mathcal{P}V(g)\mathcal{P} = \pm V(g)$.

\subsection{Classification}\label{sec:classification}
We start by considering unitary on-site symmetries, and treat the case of even and odd algebra fMPS separately.

\underline{\emph{Even algebra:}}

For an even algebra fMPS one can show using similar techniques as in the bosonic case \cite{PerezGarcia2} that the virtual symmetry matrices $V(g)$ should satisfy
\begin{align}
V(g)\mathcal{P} & = (-1)^{\mu(g)}\mathcal{P}V(g) \\
V(g)V(h) &  = \omega(g,h)V(gh)\, ,
\end{align}
with $\omega(g,h) \in \mathbb{C}$ and $\mu(g) \in \{0,1\}$ a homomorphism from $G$ to $\mathbb{Z}_2$ with the restriction that $\mu(P) = 0$. Note that $(-1)^{\mu(g)}$ is incorporated in $\omega(g,h)$ but we treat it on a separate level because it has a distinguished physical meaning, namely, it is the sign picked up by the fMPS with periodic boundary conditions under the symmetry action. Stated otherwise, on a ring with periodic boundary conditions the system transforms according the one-dimensional representation given by $\mu(g)$. It also has a distinct role in the group structure under stacking of phases as we will explain later on.

Associativity of the product of virtual symmetry matrices implies that $\omega(g,h)$ satisfies
\begin{equation}\label{cocycle}
\omega(g,h)\omega(gh,k) = \omega(g,hk)\omega(h,k)\, ,
\end{equation}
which means it is a 2-cocycle. Note that every $V(g)$ is only defined up to a complex number $\beta(g)$, implying that $\omega(g,h)$ has the same ambiguity:
\begin{equation}\label{coboundary}
\omega(g,h) \rightarrow \omega(g,h)\frac{\beta(g)\beta(h)}{\beta(gh)}\, ,
\end{equation}
where the ratio of betas appearing in the redefinition of $\omega$ is called a coboundary. So just as in the bosonic case, symmetric even algebra fMPS are classified by the different classes of $\omega(g,h)$ under the above equivalence relation. Mathematically, these classes are described by the second cohomology group $H^2(G,\mathbb{C}^*)$, or, since $H^2(G,\mathbb{R}^+)$ is trivial, by $H^2(G,U(1))$.

\underline{\emph{Odd algebra:}}

In an odd algebra fMPS the parity of $V(g)$ is ambiguous since we can always multiply $V(g)$ by $Y$, which commutes with all the fMPS tensors. We make use of this freedom to make $V(g)=V_e(g)$ even for all $g\in G$. However, for an odd fMPS with periodic boundary conditions \eqref{oddfMPS} to be invariant under the symmetry action we see that $V(g)$ has to commute or anti-commute with $Y$. This implies following properties of $V(g)$: 
\begin{align}
V(g)Y & = (-1)^{\mu(g)}YV(g) \\
V_e(g)V_e(h) & = \omega(g,h)V_e(gh)\, .
\end{align}
Here, $\mu(g) \in \{0,1\}$ is again a homomorphism from $G$ to $\mathbb{Z}_2$, but this time it is not included in $\omega$ and we have $\mu(P) = 1$. It has the same physical meaning as in the even algebra case, i.e.\ it is the one dimensional representation under which the fMPS with periodic boundary conditions transforms. $\omega(g,h) \in \mathbb{C}$ again has to satisfy equation \eqref{cocycle} and has the same ambiguity \eqref{coboundary} under multiplication of $V(g)$ by a complex number.

The fact that $\mu(P) = 1$ has a big implication. Suppose that $g^2 = P$. Then $V_e(g)^2 = \omega(g,g)\mathcal{P}$. But $\mathcal{P}Y = -Y\mathcal{P}$, which is inconsistent with the fact that $V_e(g)$ should commute or anti-commute with $Y$. In general, we can multiply all group elements with $\mu(g)=-1$ with $P$. Because $P$ commutes with everything the redefined group elements all have $\mu(g) = 0$ and form a subgroup of $G$. This implies that $G\simeq\tilde{G}\times\{I,P\}$ \cite{Fidkowski2}. So systems with particle number conservation cannot be written as an odd algebra fMPS, or physically, these systems cannot have Majorana edge modes. We give an alternative proof of this fact, based directly on the algebraic structure of the fMPS tensors, in appendix \ref{app:majorana}. We can also conclude that symmetric odd algebra fMPS are classified by $H^2(\tilde{G},U(1))$ and $H^1(\tilde{G},\mathbb{Z}_2)$, where $H^1$ denotes the homomorphism $\mu$ \cite{Fidkowski2}.

\subsection{Anti-unitary symmetries} \label{sec:antiunitary}
To incorporate anti-unitary symmetries we now introduce a new homomorphism $\gamma: G \rightarrow \mathbb{Z}_2$, where $\gamma$ denotes whether $g$ is unitary or anti-unitary. A straightforward generalization shows that the virtual symmetry matrices now have to satisfy
\begin{equation}
V(g) \lfloor V(h) \rceil^{\gamma(g)} = \omega(g,h)V(gh)\, ,
\end{equation}
where we introduced the notation
\begin{equation}
\lfloor X \rceil^{\gamma(g)} =\begin{cases}
\begin{matrix} X & \text{if } \gamma(g) = 0 \\ \bar{X} & \text{if } \gamma(g) = 1\end{matrix}\,.
\end{cases}
\end{equation}
The complex numbers, or, without loss of generality, phases $\omega(g,h)$ have the following property due to associativity
\begin{equation}\label{cocycleanti}
\omega(g,h)\omega(gh,k) = \omega(g,hk) \lfloor \omega(h,k)\rceil^{\gamma(g)}\, ,
\end{equation}
and are again ambiguous under a redefinition of $V(g)$ with scalars $\beta(g)$:
\begin{equation}\label{coboundaryanti}
\omega(g,h)\rightarrow \omega(g,h)\frac{\beta(g)\lfloor \beta(h)\rceil^{\gamma(g)}}{\beta(gh)}\, .
\end{equation}
So for non-trivial $\gamma$, symmetric fMPS are characterized by the homomorphism $\mu(g)$ and classes of $\omega(g,h)$ satisfying \eqref{cocycleanti} under the equivalence relation given in Eq.~\eqref{coboundaryanti}. One can show using similar arguments as above that, if $\mu(P)=1$, then $G \simeq \tilde{G}\times \{I,P\}$ also for non-trivial $\gamma$. This implies for example that Majorana edge states also do not appear in systems where $T^2 = P$.

We conclude the classification with a comment on the role of the phase factors $e^{i\theta(g)}$ in equations (\ref{symm1}, \ref{symm2}). These phase factors have to satisfy
\begin{equation}
e^{i\theta(g)}\lfloor e^{i\theta(h)}\rceil^{\gamma(g)} = e^{i\theta(gh)}\, .
\end{equation}
However, they are not stable under a redefinition of the unit cell, i.e. under blocking of $n$ MPS tensors the new phase factors become $e^{in\theta(g)}$. So for quantum phases that do not require strict translational symmetry on the original lattice, which is the case we concentrate on, no additional invariants can be derived from these phase factors.

\subsection{Group structure}\label{sec:groupstructure}

We now look at the group structure of phases protected by on-site symmetries under stacking. We start with the stacking of two even algebra fMPS.

\underline{\emph{Even - even:}}

If we take the graded tensor product of two even fMPS, each symmetric under the symmetry group $G$ and with respective virtual symmetry actions $V_1(g)$ and $V_2(g)$, then the virtual symmetry of the new fMPS is given by $\tilde{V}(g) = V_1(g)\otimes_{\mathfrak{g}}V_2(g)$. If a representative cocycle for $V_1(g)$ is given by $\omega_1(g,h)$ and for $V_2(g)$ by $\omega_2(g,h)$, then we see that multiplication of the new virtual symmetry action is given by
\begin{widetext}
\begin{align}
\tilde{V}(g)\lfloor \tilde{V}(h)\rceil^{\gamma(g)} & = \left(V_1(g)\otimes_{\mathfrak{g}} V_2(g)\right) \left(\lfloor V_1(h)\rceil^{\gamma(g)}\otimes_{\mathfrak{g}} \lfloor V_2(h)\rceil^{\gamma(g)}\right)  \\
& = (-1)^{\mu_1(h)\mu_2(g)}\omega_1(g,h)\omega_2(g,h)\left(V_1(gh)\otimes_{\mathfrak{g}} V_2(gh)\right)\, ,
\end{align}
where $\mu_{1(2)}(g)$ denotes the parity of $V_{1(2)}(g)$. 
\end{widetext}
So the symmetry-protected phase of the stacked fMPS is captured by following algebraic data:
\begin{align}
\tilde{\mu}_{ee}(g) & = \mu_1(g) + \mu_2(g) \text{ mod } 2 \\ 
\tilde{\omega}_{ee}(g,h) & = (-1)^{\mu_1(h)\mu_2(g)}\omega_1(g,h)\omega_2(g,h)\, .
\end{align}
Note that this expression is symmetric since $\mu_1(h)\mu_2(g) + \mu_2(h)\mu_1(g) = \mu_1(gh)\mu_2(gh) + \mu_1(g)\mu_2(g) + \mu_1(h)\mu_2(h)$ can be absorbed by a redefinition of the virtual symmetry actions.

\underline{\emph{Even - odd:}}

To expose the behaviour of the invariants under stacking of an even and odd algebra fMPS we first note that the virtual symmetry action for an odd algebra fMPS takes the form
\begin{align}
V_e(g)& = \mathds{1}\otimes V'(g)\;\;\; \text{ if }\;\;\;\; \mu(g) = 0\\
V_e(g)& = z\otimes V'(g) \;\;\;\text{ if }\;\;\;\; \mu(g) = 1\, .
\end{align}
The virtual symmetry of the stacked fMPS is again the graded tensor product of $V_{1e}(g) = z^{\mu_1(g)}\otimes V_1'(g)$ and $V_2(g)$, corresponding to the odd and even algebra fMPS respectively. This graded tensor product takes on the parity of $V_2(g)$. However, the stacked fMPS is again of odd algebra type so to expose the invariants all virtual symmetry actions should be even. To accomplish this we multiply all $V_{1e}(g)\otimes_{\mathfrak{g}}V_2(g)$ for which $\mu_2(g) = 1$ with $Y\otimes_{\mathfrak{g}}\mathds{1}$, the odd central element of the tensors of the stacked fMPS. This leads to the following expression for the even virtual symmetry actions of the stacked fMPS:
\begin{equation}
\tilde{V}_e(g) = z^{\mu_1(g)}y^{\mu_2(g)}\otimes V'_1(g)\otimes_{\mathfrak{g}} V_2(g)
\end{equation}
We can now extract the new $\tilde{\mu}$ invariant 
\begin{align}
\tilde{V}_e(g)Y &= \left(z^{\mu_1(g)}y^{\mu_2(g)}\otimes V'_1(g)\otimes_{\mathfrak{g}} V_2(g)\right)(y\otimes\mathds{1}\otimes_{\mathfrak{g}}\mathds{1})\nonumber \\
 & = (-1)^{\mu_2(g)}z^{\mu_1(g)}y^{\mu_2(g)}y\otimes V'_1(g)\otimes_{\mathfrak{g}} V_2(g)\nonumber \\
 & = (-1)^{\mu_1(g)+\mu_2(g)}yz^{\mu_1(g)}y^{\mu_2(g)}\otimes V'_1(g)\otimes_{\mathfrak{g}} V_2(g)\nonumber \\
& = (-1)^{\mu_1(g)+\mu_2(g)}Y\tilde{V}_e(g)
\end{align}
\begin{widetext}
In a similar way we can obtain a representative 2-cocycle for the stacked fMPS:
\begin{align}
\tilde{V}_e(g)\tilde{V}_e(h) & = \left( z^{\mu_1(g)}y^{\mu_2(g)}\otimes V'_1(g)\otimes_{\mathfrak{g}} V_2(g)\right)\left(z^{\mu_1(h)}y^{\mu_2(h)}\otimes \lfloor V'_1(h)\rceil^{\gamma(g)}\otimes_{\mathfrak{g}} \lfloor V_2(h)\rceil^{\gamma(g)}\right)\\
 & = (-1)^{\mu_2(h)\mu_2(g)}z^{\mu_1(g)}y^{\mu_2(g)}z^{\mu_1(h)}y^{\mu_2(h)}\otimes V'_1(g)\lfloor V'_1(h)\rceil^{\gamma(g)}\otimes_{\mathfrak{g}}V'_2(g)\lfloor V'_2(h)\rceil^{\gamma(g)}\\
& = (-1)^{\mu_2(g)(\mu_1(h)+\mu_2(h))}\omega_1(g,h)\omega_2(g,h) \\ 
& \;\;\;\;\;\;\;\; z^{\mu_1(g) + \mu_1(h)}y^{\mu_2(g)+\mu_2(h)}\otimes V'_1(gh)\otimes_{\mathfrak{g}}V'_2(gh)
\end{align}
Using the fact that $\mu_1$ and $\mu_2$ are homomorphisms, and that $y^2 = -\mathds{1}$ we find
\begin{align}
\tilde{V}_e(g)\tilde{V}_e(h) = & (-1)^{\mu_2(g)(\mu_1(h)+\mu_2(h))}\omega_1(g,h)\omega_2(g,h)i^{\mu_2(g) + \mu_2(h) - \mu_2(gh)} \\
 & \;\;\;\; z^{\mu_1(gh)}y^{\mu_2(gh)}\otimes V'_1(gh)\otimes_{\mathfrak{g}}V'_2(gh)
\end{align}
\end{widetext}
So, putting everything together, we conclude that the stacked odd algebra fMPS has invariants described by the data
\begin{align}
\tilde{\mu}_{eo}(g) & = \mu_e(g) + \mu_o(g) \text{ mod } 2\\
\tilde{\omega}_{eo}(g,h) & = (-1)^{\mu_e(g) \mu_o(h)}\omega_e(g,h)\omega_o(g,h)\, ,
\end{align}
where we replaced the subscripts $1$ and $2$ with $o$ and $e$ to denote the odd and even fMPS that are being stacked.  

\underline{\emph{Odd - odd:}}

Before we derive the behaviour of the invariants in symmetric odd algebra fMPS under stacking we first make a few observations. Since $U(g)P = PU(g)$ we have $U(g) = \left(\begin{matrix} U^0(g) & 0 \\ 0 & U^1(g) \end{matrix}\right)$ and we can write the local condition for the fMPS to be invariant under the symmetry as
\begin{widetext}
\begin{align}
\sum_{j:|j| = |i|} \left(U^{|i|}(g)\right)_{ij}\left(y^{|j|}\otimes \lfloor B^j \rceil^{\gamma(g)}\right) & = \left(z^{\mu(g)}\otimes V'(g)^{-1} \right) \left(y^{|i|}\otimes B^i \right) \left( z^{\mu(g)}\otimes V'(g)\right) \\
 & = (-1)^{|i|\mu(g)}\left(\mathds{1}\otimes V'(g)^{-1} \right) \left(y^{|i|}\otimes B^i \right) \left( \mathds{1}\otimes V'(g)\right)
\end{align}

From this we see that

\begin{equation}\label{localodd}
\sum_{j:|j| = |i|} \left(U^{|i|}(g)\right)_{ij} \lfloor B^j\rceil^{\gamma(g)} = (-1)^{|i|\mu(g)}V'(g)^{-1} B^i V'(g)
\end{equation}
In section \ref{Z2group} we also learned that the fMPS tensors for the stacked state take the form

\begin{equation}
\begin{array}{ll}
\tilde{A}^{ij} = \mathds{1} \otimes B^i_1 \otimes B^j_2  & \text{if }  |i| = |j| = 0\\
\tilde{A}^{ij} = -i x\otimes B^i_1\otimes B^j_2  &\text{if }  |i| = 0  \text{ and }  |j| = 1 \\
\tilde{A}^{ij} = y\otimes B^i_1\otimes B^j_2  &\text{if }   |i| = 1  \text{ and }  |j| = 0  \\
\tilde{A}^{ij} = -iz\otimes B^i_1\otimes B^j_2  &\text{if }   |i| = |j| = 1\, .
\end{array}
\end{equation}
Combining this expression for the fMPS tensors with equation \eqref{localodd} one sees that the virtual symmetry actions of the stacked even algebra fMPS are

\begin{equation}
\tilde{V}(g) = x^{\mu_1(g)}y^{[\mu_2(g)+\gamma(g)]}\otimes V_1'(g)\otimes V'_2(g)\, ,
\end{equation}
where $[\cdot] \in \{0,1\}$ denotes modulo 2. The parity of this matrix is $\mu_1(g) + \mu_2(g)+\gamma(g)$ mod 2. 

Multiplication of these virtual symmetry actions also gives us a representative 2-cocycle and we see that the invariants of the stacked fMPS are described by

\begin{align}
\tilde{\mu}_{oo}(g) & = \mu_1(g) + \mu_2(g) + \gamma(g) \text{ mod } 2 \\ 
\tilde{\omega}_{oo}(g,h) & = (-1)^{\mu_1(h)(\mu_2(g)+\gamma(g))}i^{[\mu_2(g)+\gamma(g)] + [\mu_2(h)+\gamma(h)] - [\mu_2(gh)+\gamma(gh)]}\omega_1(g,h)\omega_2(g,h)
\end{align}
\end{widetext}
We note that this expression at first sight does not look symmetric in $\mu_1$ and $\mu_2$. However, the symmetry can be understood from the fact that the stacked fMPS tensors of $|\psi\rangle_1\otimes_{\mathfrak{g}}|\psi\rangle_2$ and  $|\psi\rangle_2\otimes_{\mathfrak{g}}|\psi\rangle_1$ are related by a unitary gauge tranformation $U=u\otimes\mathds{1}\otimes\mathds{1}$, where u transforms $-ix$ into $y$ and vice versa. The virtual symmetry matrices then transform as $U\tilde{V}(g)U^\dagger$, if $\gamma(g) = 0$ and as $U\tilde{V}(g)U^T$ if $\gamma(g)=1$, effectively interchanging $\mu_1$ and $\mu_2$ and possibly adding some phase factors to the $\tilde{V}(g)$.

\section{Reflection symmetry}\label{sec:reflection}

Up to now we have studied the classification and group structure of on-site unitary and anti-unitary symmetries. In this last section we study a spatial symmetry, namely reflection symmetry. To perform a spatial reflection we need to contract the tensors in a different order. For this we first reorder the left and right virtual modes in a single fMPS tensor, resulting in:
\begin{align}\label{reflstep1}
\sum_{i\alpha\beta} A^i_{\alpha\beta}|\alpha)|i\rangle (\beta| \rightarrow \sum_{i\alpha\beta} A^i_{\alpha\beta}(-1)^{|i||\alpha| + |\beta|}(\beta|\, |i\rangle  |\alpha)
\end{align}
Because the roles of virtual bras and kets are switched, the contraction of the virtual bonds in the fMPS is of the form $\mathcal{C}(|\alpha) \otimesg (\alpha|)$, i.e.\ a supertrace, and an extra factor $\mathcal{P}$ is picked up on every bond. Alternatively, we can interchange the bras and the kets between subsequent fMPS tensors in order to recover normal fMPS tensors where the left virtual mode is a ket and the right virtual mode is a bra. This yields the same sign factor, and we thus obtain:
\begin{equation}\label{reflstep2}
\sum_{i\alpha\beta} A^i_{\alpha\beta}(-1)^{|i||\alpha|}|\beta)|i\rangle  (\alpha| = \sum_{i\alpha\beta} A^i_{\beta\alpha}(-1)^{|i||\beta|}|\alpha)|i\rangle (\beta|
\end{equation}
Apart from this change in contraction order, reflection can additionally involve an onsite unitary $U_R$, with $U_RP =PU_R$. If the transformed fMPS tensors represents the same state then we know they should be related to the original tensors via a gauge transformation $\mathcal{R}$. So the local condition for a reflection symmetric fMPS becomes
\begin{equation}\label{reflection}
\sum_{j}\left(U_R\right)_{ij} \left(A^j\right)^T\mathcal{P}^{|j|}  = \mathcal{R}A^i\mathcal{R}^{-1}
\end{equation}
In the remainder of this section we will study the situation where $U_R^2 = P$. The case $U_R^2 = \mathds{1}$ can be worked out similarly. If we apply reflection twice then we get following condition:
\begin{equation}\label{reflectiontwice1}
P_{ii}A^i = (-1)^{|i|\mu(R)}\mathcal{P}^{|i|}(\mathcal{R}^{-1\,T}\mathcal{R})A^i(\mathcal{R}^{-1}\mathcal{R}^T)\mathcal{P}^{|i|}\, ,
\end{equation}
where $\mu(R) \in \{0,1\}$ denotes the parity of $\mathcal{R}$, i.e. $\mathcal{R}\mathcal{P} = (-1)^{\mu(R)}\mathcal{P}\mathcal{R}$. Using $P_{ii} = (-1)^{|i|}$ and $\mathcal{P}^{|i|} A^i \mathcal{P}^{|i|} = \mathcal{P} A^i \mathcal{P} = (-1)^{|i|} A^i$, Eq.~\eqref{reflectiontwice1} can be rewritten as
\begin{equation}\label{reflectiontwice2}
P_{ii}^{\mu(R)}A^i =(\mathcal{R}^{-1\,T}\mathcal{R})A^i(\mathcal{R}^{-1}\mathcal{R}^T)\, .
\end{equation}

\subsection{Classification}
To extract the discrete invariants that label different reflection symmetric phases we again consider the cases of even and odd graded algebras separately.

\underline{\emph{Even algebra:}}

If $\mu(R) = 0$ then we see from equation \eqref{reflectiontwice2} that
\begin{equation}
\mathcal{R} = \alpha \mathcal{R}^T\, ,
\end{equation}
with $\alpha \in \mathbb{C}$. However, by taking the transpose of this expression one learns that $\alpha^2 = 1$. So we obtain the invariant $\rho \in \{0,1\}$, which denotes whether $\mathcal{R}$ is symmetric or anti-symmetric:
\begin{equation}
\mathcal{R} = (-1)^\rho \mathcal{R}^T\, .
\end{equation}
If $\mu(R) = 1$ then equation \eqref{reflectiontwice2} implies that
\begin{equation}
\mathcal{R} = \alpha \mathcal{R}^T\mathcal{P}\, .
\end{equation}
Again taking the transpose of this expression we can obtain a similar condition on $\mathcal{R}$:
\begin{equation}
\mathcal{R} = -\alpha^{-1} \mathcal{R}^T\mathcal{P}\, .
\end{equation}
So for $\mu(R) = 1$ the invariant $\rho$ carries the following information about $\mathcal{R}$:
\begin{equation}
\mathcal{R} = (-1)^\rho i\mathcal{R}^T\mathcal{P}\, .
\end{equation}
In total we thus obtain four possible reflection symmetric phases in even algebra fMPS, labeled by the invariants $\mu(R)$ and $\rho$.

\underline{\emph{Odd algebra:}}

For an odd algebra fMPS we can choose $\mathcal{R} \equiv \mathcal{R}_e$ in equation \eqref{reflection} to be even, i.e. $\mu(R) = 0$. Just as in the even algebra case this gives rise to an invariant $\rho_1 \in \{0,1\}$:
\begin{equation}
\mathcal{R}_e = (-1)^{\rho_1}\mathcal{R}^T_e\, .
\end{equation}
However, $\mathcal{R}_o \equiv \mathcal{R}_eY$ can also serve as $\mathcal{R}$ in equation \eqref{reflection}. Because $\mathcal{R}_o$ is odd, we can define a second invariant $\rho_2 \in \{0,1\}$ by doing the same steps as in the even algebra case above:
\begin{equation}
\mathcal{R}_o = (-1)^{\rho_2} i\mathcal{R}_o^T\mathcal{P}
\end{equation}
One can check that the conditions on $\mathcal{R}_e$ and $\mathcal{R}_o$ are independent and imply that $\mathcal{R}_e$ is of the form
\begin{equation}
\mathcal{R}_e = \left(\begin{matrix}(-1)^{\rho_1 + \rho_2}i & 0 \\ 0 & 1 \end{matrix}\right)\otimes \mathcal{R}'\;\;\;\text{ with }\;\;\; \mathcal{R}'^T = (-1)^{\rho_1}\mathcal{R}'\, .
\end{equation}
We thus again obtain four possibilities labeled by $\rho_1$ and $\rho_2$, leading to a total of 8 different reflection symmetric phases. Note that $\mathcal{R}_e$ does not commute or anti-commute with $Y$. This is also not required because reflection symmetry cannot be defined on a chain with periodic boundary conditions. If we would have used a different convention to define the odd gauge transformation, i.e. $\mathcal{R}_o' \equiv Y\mathcal{R}_e$, then we would obtain an equivalent invariant $\rho_2'$, which is related to $\rho_2$ via $\rho_2' = \rho_2 + 1$ mod 2.

\subsection{Partial reflection}

To expose the group structure of the eight reflection symmetric phases we will use a different approach than in section \ref{sec:onsite}. It was argued in Refs.~\cite{Hsieh1,Hsieh2,Kapustin1,Kapustin2,Witten1,Witten2,Metlitski} that the phase of the partition function of reflection symmetric phases on an unorientable spacetime is an invariant. It was shown in Refs.~\cite{Shapourian,Shiozaki} that this invariant phase on $\mathbb{R}P^2$ can be obtained from the ground state wave function. The approach is similar to the bosonic case \cite{Cen,Pollmann}, and is based on the calculation of the ground state expectation value of a non-local operator $R_{\text{part}}$, called partial reflection.

We focus on the Kitaev chain described by Hamiltonian \eqref{KitaevChain}. This Hamiltonian has reflection symmetry given by
\begin{equation}
a_j \rightarrow i a_{-j}
\end{equation}
So we see that
\begin{equation}
U_{R} = \left(\begin{matrix}1 & 0 \\ 0 & i \end{matrix}\right)\, ,
\end{equation}
and $U_{R}^2 = P$. The MPS tensors of the Kitaev chain ground state satisfy Eq.~\eqref{reflection}, with
\begin{equation}
\mathcal{R}\equiv \mathcal{R}_e = \left(\begin{matrix}e^{i\pi/4} & 0 \\ 0 & e^{-i\pi/4}  \end{matrix}\right)
\end{equation}
To define the partial reflection operator we divide up the chain with periodic boundary conditions into two connected intervals $I_1$ and $I_2$ of length $N_1$ and $N_2$. Partial reflection then acts as normal reflection, but only on one of the two intervals, which we take to be $I_1$.

To apply $R_{\text{part}}$ in the fMPS formalism we first perform step \eqref{reflstep1} and \eqref{reflstep2} on the MPS tensors in $I_1$, together with the on-site unitary $U_R$. Note that we can only interchange virtual kets and bras between subsequent tensors [i.e. Eq.~\eqref{reflstep2}] in the bulk of $I_1$ but not at the end points. Using Eq.~\eqref{reflection} this gives the following concatenated tensor corresponding to $I_1$:
\begin{equation}
\sum_{i_1,\dots,i_{N_1}}\left(\mathcal{R}_eA^{i_{N_1}}\dots A^{i_1}\mathcal{R}_e^{-1} \right)_{\alpha\beta}(-1)^{|\alpha|}(\alpha|\,|i_{N_1}\rangle\dots |i_1\rangle |\beta)\, .
\end{equation}
\begin{widetext}
For $I_2$ we have the usual concatenated tensor
\begin{equation}
\sum_{i_{N_1+1},\dots,i_{N_1+N_2}}\left( A^{i_{N_1+1}}\dots A^{i_{N_1+N_2}}y\right)_{\alpha\beta}|\alpha)|i_{N_1+1}\rangle\dots |i_{N_1+N_2}\rangle (\beta|\, .
\end{equation}
To obtain the total wavefunction we have to contract the virtual $\alpha$ and $\beta$ indices. The final expression for $R_{\text{part}}|\psi\rangle$ is then 
\begin{align}
R_{\text{part}}|\psi\rangle = \sum_{i_1\dots i_{N_1+N_2}}\sum_{\alpha\beta}\left(\mathcal{R}_eA^{i_{N_1}}\dots A^{i_1}\mathcal{R}_e^{-1} \right)_{\alpha\beta}\left( A^{i_{N_1+1}}\dots A^{i_{N_1+N_2}}y\right)_{\alpha\beta} 
(-1)^{|\beta|(|\alpha|+|\beta|)}|i_{N_1}\rangle\dots|i_1\rangle|i_{N_1+1}\rangle\dots|i_{N_1+N_2}\rangle
\end{align}
To calculate the expectation value we note that $\langle\psi|$ is given by
\begin{equation}
\langle\psi| = \sum_{i_1\dots i_{N_1+N_2}} \text{tr}\left(\bar{A}^{i_{N_1}}\dots \bar{A}^{i_1}\bar{A}^{i_{N_1+1}}\dots \bar{A}^{i_{N_1+N_2}}y\right)\langle i_{N_1+N_2}|\dots\langle i_{N_1+1}|\langle i_1|\dots\langle i_{N_1}|
\end{equation}
\end{widetext}
We introduce following graphical notation for the transfer matrix (note that without arrows, the graphical notation denotes the tensor components):
\begin{equation}
\includegraphics[width=0.8\linewidth]{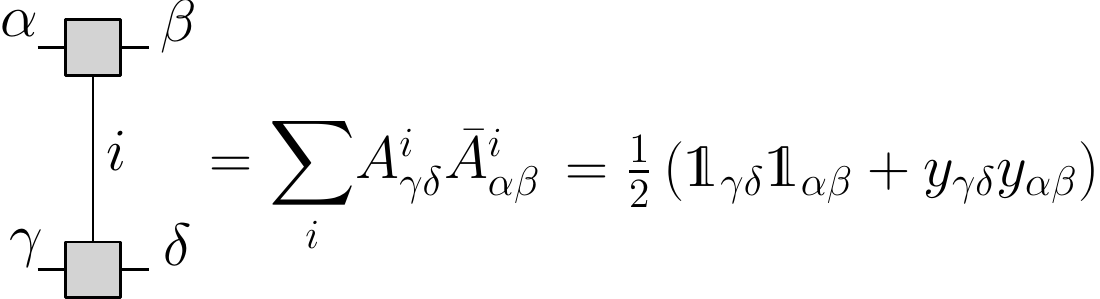}\, ,
\end{equation}
where the normalization factor $1/2$ has been inserted to ensure that the transfer matrix is a projector (which is the fMPS manifestation of the fixed point character of the model). With this notation we can represent $\langle\psi|R_{\text{part}}|\psi\rangle$ as
\begin{equation}
\includegraphics[width=0.95\linewidth]{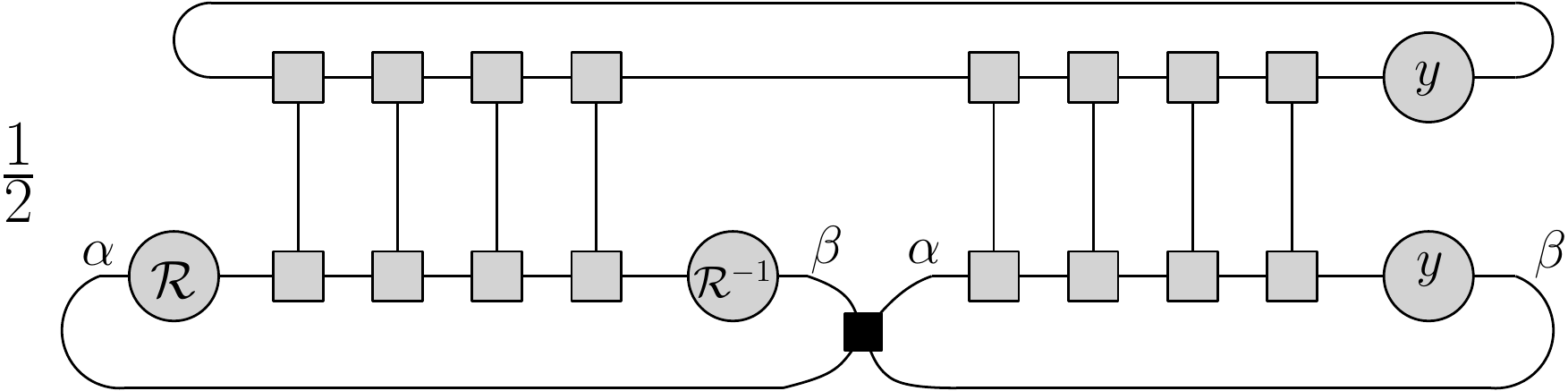}
\end{equation}
The global factor $1/2$ comes from the normalization of $|\psi\rangle$ and the black square denotes the non-local sign $(-1)^{|\beta|(|\alpha|+|\beta|)}$. 
\begin{widetext}
Putting everything together we easily obtain following result
\begin{equation}
\langle\psi |R_{part}|\psi\rangle = \frac{1}{8}\left(\text{tr}(yy)\text{tr}(\mathcal{R}\mathcal{R}^{-1}y^Ty^T)+\text{tr}(yy)\text{tr}(\mathcal{R}y\mathcal{R}^{-1}zy^T) \right)=\frac{1+i}{2} = \frac{e^{i\pi/4}}{\sqrt{2}}\, .
\end{equation}
\end{widetext}
This agrees with the previous studies of reflection symmetric phases on $\mathbb{R}P^2$. We see that in order to obtain a trivial phase factor one needs eight copies of the Majorana chain, which shows that reflection symmetric phases in one spatial dimension form a $\mathbb{Z}_8$ group.

\section{Discussion}

We have shown that the formalism of fermionic matrix product states captures all phases of interacting fermions in one dimension, both for general on-site and spatial symmetries. All universal information about the quantum phase, including the presence or absence of Majorana zero modes, can be extracted in a local fashion from the tensor building up the ground state wave function. This local encoding of the information allows for a straightforward analysis of the group structure of symmetric phases under stacking. An advantage of the framework is that it provides a physical interpretation for the invariants labeling the different phases, both in terms of the entanglement spectrum and quantum numbers of the ground state without or with a background gauge field. The latter case can be studied analogously to the bosonic case \cite{Zaletel}. 
\\ \\
Fermionic tensor network states, and particularly fermionic MPOs, can also be used to describe topological phases in higher dimensions  \cite{Schuch3,Wahl,Dubail,Gu2,Gu3,Gu4,Gaiotto,Bhardwaj,Tarantino,Ware,Williamson2,Wille}. In this way the observations made in this work should be relevant for a systematic understanding of how universal topological properties of two and three-dimensional systems, such as the binding of Majorana modes to symmetry defects, are connected to the entanglement structure of the ground state wave function. We hope that this work may pave the way to a more detailed understanding of higher dimensional topological fermionic tensor networks. 
\\ \\
\emph{Acknowledgements - } This work was supported by the Austrian Science Fund (FWF) through grants ViCoM and FoQuS, and the EC through the grant QUTE. J.H. and F.V. acknowledge the support from the Research Foundation Flanders (FWO).

\vspace{10 mm}
\appendix

\section{Translation symmetry for fermionic systems}\label{app:translation}

At the single particle level, the translation operator $T_P$ with periodic boundary conditions is defined as
\begin{equation}
T_Pa_jT_P^{-1} = a_{j+1}
\end{equation}
Translation with a $\pi$-flux inserted through the system, or equivalently, with anti-periodic boundary conditions is defined as
\begin{align}
T_{AP}a_jT_{AP}^{-1} & = a_{j+1}\;\;\;\;\text{ for }\;\; j\neq N\\
T_{AP}a_NT_{AP}^{-1} & = -a_{1}\, ,
\end{align}
where $N$ is the number of sites in the system. One obviously has
\begin{align}
T_P^N &= \mathds{1}\\
T_{AP}^N &= -\mathds{1}\, ,
\end{align}
so the eigenvalues $e^{ik}$ of $T_P$ have momenta $k = \frac{2\pi}{N}n$ with $n\in\{0,1,\dots,N-1\}$ while those of $T_{AP}$ have momenta $k = \frac{\pi}{N}(2n+1)$ with $n \in \{0,1,\dots , N-1\}$. We would now like to implement $T_P$ and $T_{AP}$ on states in the fermionic many-body Hilbert space. 
\begin{widetext}
We start with a general even state $|\psi_e\rangle$:
\begin{equation}
|\psi\rangle_e = \sum_{\{i\}=0}^1 \psi^e_{i_1i_2\dots i_N}|i_1\rangle_1|i_2\rangle_{2}\dots|i_N\rangle_N \in \mathcal{H}_1\otimes \mathcal{H}_2\otimes\dots\otimes \mathcal{H}_N\, ,
\end{equation}
where $\sum_{j=1}^N |i_j| = 0$ mod 2. Applying $T_P$ on $|\psi\rangle_e$ gives following state:
\begin{equation}
T_P|\psi\rangle_{e}  =  \sum_{\{i\}=0}^1 \psi^e_{i_1i_2\dots i_N}|i_1\rangle_2|i_2\rangle_{3}\dots|i_N\rangle_1 \in \mathcal{H}_2\otimes \mathcal{H}_3\otimes\dots\otimes \mathcal{H}_1
\end{equation}
To compare $T_P|\psi\rangle_e$ with $|\psi\rangle_e$ with need to use the fermionic tensor product isomorphism to map the translated state into the same Hilbert space as the original state. Using the fact that $\sum_{j=1}^{N-1}|i_j| = |i_N|$ mod 2 we find
\begin{eqnarray}
 \mathcal{F}\left(T_P|\psi\rangle_e \right) & = & \sum_{\{i\}=0}^1 \psi^e_{i_1i_2\dots i_{N}} (-1)^{|i_N|}|i_N\rangle_1|i_1\rangle_{2}\dots|i_{N-1}\rangle_N \in \mathcal{H}_1\otimes \mathcal{H}_2\otimes\dots\otimes \mathcal{H}_N \nonumber\\
 & = & \sum_{\{i\}=0}^1 \psi^e_{i_2i_3\dots i_1}(-1)^{|i_1|}|i_1\rangle_1|i_2\rangle_{2}\dots|i_N\rangle_N
\end{eqnarray}
\end{widetext}
So the condition that $|\psi\rangle_e$ is an eigenstate of $T_P$ becomes
\begin{equation}
 \mathcal{F}\left(T_P|\psi\rangle_e \right) = e^{ik}|\psi\rangle_e \Rightarrow \psi^e_{i_1i_2\dots i_N} = e^{-ik} \psi^e_{i_2i_3\dots i_1}(-1)^{|i_1|}
\end{equation}
Similarly, one can check that the condition for $|\psi\rangle_e$ to be an eigenstate of $T_{AP}$ is
\begin{equation}
\mathcal{F}\left(T_{AP}|\psi\rangle_e \right) = e^{ik}|\psi\rangle_e \Rightarrow \psi^e_{i_1i_2\dots i_N} = e^{-ik} \psi^e_{i_2i_3\dots i_1}
\end{equation}
For odd states $|\psi\rangle_o$, i.e. states with $\sum_{j=1}^N |i_j| = 1$ mod 2, we find 
\begin{align}
 \mathcal{F}\left(T_P|\psi\rangle_o \right) &= e^{ik}|\psi\rangle_o \Rightarrow \psi^o_{i_1i_2\dots i_N} = e^{-ik} \psi^e_{i_2i_3\dots i_1} \\
 \mathcal{F}\left(T_{AP}|\psi\rangle_o \right) &= e^{ik}|\psi\rangle_o \Rightarrow \psi^o_{i_1i_2\dots i_N} = e^{-ik} \psi^o_{i_2i_3\dots i_1}(-1)^{|i_1|}
\end{align}
We close this appendix by noting that for the classification of topological phases we consider in the main text, strict translational symmetry is not required. However, if one was to incorporate translational symmetry in the classification the number of phases would (at least) double because the fMPS tensors can be even or odd. We always assumed the fMPS tensors to be even, which can be done by a redefinition of the unit cell and blocking two tensors.

\section{Parity of gauge transformations}\label{app:parity}
Gauge transformations correspond to different choices of basis in the virtual spaces of the tensor network. For fermionic tensor networks where the virtual spaces also have a $\mathbb{Z}_2$ grading, it is only sensible to work with a basis where the grading is explicit, and thus where $\mathcal{P}$ takes the standard form $\mathds{1}_{D_e}\oplus \mathds{1}_{D_o}$ before and after the transformation. In particular, we require that $A^i$ and its gauge transformed version $A'^i=M A^i M^{-1}$ are even tensors with respect to the same $\mathcal{P}$, so we have that
\begin{equation}
\mathcal{P}MA^iM^{-1}\mathcal{P} = M\mathcal{P}A^i\mathcal{P}M^{-1}
\end{equation}
This relation implies that
\begin{equation}
\mathcal{P}M = M\mathcal{P}X^{-1}\, ,
\end{equation}
where $X$ is an invertible matrix in the center of the algebra spanned by the tensors $A^i$.

If the graded algebra is of even type this implies that 
\begin{equation}
\mathcal{P}M = \pm M\mathcal{P}\, ,
\end{equation}
so the gauge transformation $M$ has a well-defined parity. Note that an odd invertible matrix $M$ (which interchanges the even and the odd basis vectors of the virtual space) can only exist if $D_e = D_o$.

If $A^i$ span an odd algebra then we have
\begin{equation}
\mathcal{P}M = M\mathcal{P}(\alpha\mathds{1} + \beta Y)^{-1}\, ,
\end{equation}
with $\alpha \neq \pm i\beta$. We split up $M$ into its even and odd part, $M= M_e +M_o$, and obtain from the equation above
\begin{align} \label{parity1}
(1-\alpha)M_e &= +\beta M_oY \\
(1+\alpha)M_o &= -\beta M_eY \label{parity2}
\end{align}
We first consider some special cases. If $\alpha = 1$, then the above equations imply that $\beta = 0$ and $M_o = 0$. If $\alpha = -1$, then $\beta = 0$ and $M_e = 0$. So in both cases $M$ has a well-defined parity. We exclude $\alpha = \pm 1$ and $\beta = 0$ in the following steps. From equations \eqref{parity1} and \eqref{parity2} we find that
\begin{align}
M_e = \frac{\beta}{1-\alpha}M_oY = \frac{1+\alpha}{\beta}M_oY\, ,
\end{align}
which implies that $\alpha^2 + \beta^2 = 1$. We can now write $M$ as
\begin{equation}
M = M_o\left(\mathds{1}\pm\sqrt{\frac{1+\alpha}{1-\alpha}}Y \right) \equiv M_oX\, ,
\end{equation}
where $X$ is invertible. Because $X$ is in the center of the odd algebra spanned by $A^i$, the gauge transformation $M_o$ also relates $A'^i$ to $A^i$:
\begin{equation}
A'^i = M_oA^iM_o^{-1}\, .
\end{equation}
This shows that also in the case where $A^i$ span an odd algebra we can without loss of generality restrict to gauge transformations that have a well-defined parity.

\section{Majorana modes and superconductivity}\label{app:majorana}
In this appendix we show that the structure of the tensors in an odd algebra fMPS is incompatible with a $U(1)$ charge conservation symmetry. This implies that Majorana edge modes can only appear in superconductors. The local condition for the fMPS to be symmetric under the global $U(1)$ symmetry is
\begin{equation}
\sum_{j}U(\theta)_{ij}A^{j} = e^{ip\theta}\,V(\theta)A^i V(\theta)^\dagger\, ,
\end{equation}
with $p \in \mathbb{Z}$. We also assume that the fMPS is irreducible. If we write $U(\theta) = \text{exp}(iq\theta)$ and $V(\theta) = \text{exp}(iQ\theta)$ and take the derivative of the above equation with respect to $\theta$ evaluated at $\theta = 0$, then we get
\begin{equation}\label{localu1}
\sum_{j}\left(q-p\mathds{1} \right)_{ij}A^{j} = [Q, A^i]\, .
\end{equation}
We continue to work in the basis for the local physical Hilbert space in which $q$ is diagonal. Because $U(1)$ corresponds to charge conservation, $q$ is of the form
\begin{equation}
q = \left(\begin{matrix} e_1 &  & & & & & \\  &\ddots & & & & &  \\
 &  & e_r & & & & \\  &  & &o_1 & & &  \\  &  & & & & \ddots&  \\  &  & & & & & o_s
\end{matrix} \right) \;\;\;\;\text{with } r+s =  d\, ,
\end{equation}
where $d$ is the dimension of the local Hilbert space and $e_i$, the eigenvalues of states with even fermion parity, are even integers, while $o_i$ are odd integers corresponding to eigenstates with odd fermion parity. Without loss of generality, we take $Q$ to be an even matrix:
\begin{equation}
Q = \left( \begin{matrix} Q_1 & 0 \\ 0 & Q_2 \end{matrix}\right)
\end{equation}
We recall that the tensors of an odd algebra fMPS take the form 
\begin{align}
\left(\begin{matrix}B^i& 0 \\ 0 & B^i \end{matrix}\right) \;\;\;\;\;\text{ if }\;\;\;\;\;|i| = 0\\
\left(\begin{matrix}0& C^i \\ -C^i & 0\end{matrix}\right) \;\;\;\;\;\text{ if }\;\;\;\;\;|i| = 1
\end{align}
From this we see that equation \eqref{localu1} is equivalent to 
\begin{eqnarray}\label{conditionQ}
[Q_1,B^i] & = & [Q_2,B^i] = (e_i - p)B^i \nonumber\\
\{Q_1,C^i\} & = & \{Q_2,C^i\} \nonumber \\
Q_1C^i - C^iQ_2 & = & (o_i-p)C^i\, ,
\end{eqnarray}
which implies that $[Q_1-Q_2,B^i] = \{Q_1-Q_2,C^i\} = 0$. However, because of the irreducibility of the fMPS we know that products of $B^i$ and $C^i$ span a full $D/2\times D/2$ matrix algebra, where $D$ is the bond dimension of the fMPS. This allows us to conclude that $Q_1 = Q_2$. Therefore, equations \eqref{conditionQ} reduce to
\begin{align}
[Q_1,B^i] = (e_i - p)B^i \nonumber \\
[Q_1,C^i] = (o_i - p)C^i \label{commutatorQ}
\end{align}
We now work in the basis in which $Q_1$ takes following diagonal form
\begin{equation}
Q_1 = \left(\begin{matrix} \lambda_{e_1} &  & & & & & \\  &\ddots & & & & &  \\
 &  & \lambda_{e_t} & & & & \\  &  & &\lambda_{o_1} & & &  \\  &  & & & & \ddots&  \\  &  & & & & & \lambda_{o_u}
\end{matrix} \right)\;\;\;\;\;\text{with }t+u = D/2\, ,
\end{equation}
where $\lambda_{e_i} \in 2\mathbb{Z}$ and $\lambda_{o_i} \in 2\mathbb{Z} + 1$. In this basis, equations \eqref{commutatorQ} can be written as
\begin{eqnarray}
(\lambda_j - \lambda_k)B^i_{jk} = (e_i - p)B^i_{jk}\\
(\lambda_j - \lambda_k)C^i_{jk} = (o_i - p)C^i_{jk}
\end{eqnarray}
If $p$ is even, this implies that the $B^i$ are block diagonal and the $C^i$ are block off-diagonal. If $p$ is odd, then the $B^i$ are block off-diagonal and the $C^i$ are block diagonal. The situations with $p$ odd and $p$ even clearly become equivalent after blocking two tensors. However, this structure of the $B^i$ and $C^i$ is in contradiction with the irreducibility of the fMPS, which requires the even subalgebra spanned by the fMPS matrices to be simple (see section \ref{sec:irreducibility}). If the fMPS were not irreducible we could write it as a sum of multiple irreducible fMPS, each of which should have $U(1)$ charge symmetry (because $U(1)$ is continuous and connected it cannot permute the different irreducible fMPS), thus again leading to a contradiction. This shows that fMPS with an odd algebra structure cannot have charge conservation. Since fMPS represent the ground state of gapped local Hamiltonians, this does not exclude the possibility of having Majorana edge modes in gapless systems with particle number conservation. Indeed, explicit examples of such systems have been constructed in the literature \cite{Fidkowski1,Cheng,Tsvelik,Sau}.

\bibliography{FermionicMPS}

\end{document}